\documentclass[aps,pra,showpacs,superscriptaddress,twocolumn]{revtex4}
\usepackage{amsmath}
\usepackage{amsfonts}
\usepackage{graphicx}
\usepackage{float}
\usepackage{color}

\begin{document}
\title{A quantum optical description of photon statistics and cross-correlations in high harmonic generation}
\author{\'{A}kos Gombk\"{o}t\H{o}}
\affiliation{Department of Theoretical Physics, University of Szeged, Tisza Lajos k\"{o}r%
	\'{u}t 84, H-6720 Szeged, Hungary}
\author{P\'{e}ter F\"{o}ldi}
\affiliation{Department of Theoretical Physics, University of Szeged, Tisza Lajos k\"{o}r%
	\'{u}t 84, H-6720 Szeged, Hungary}
\affiliation{ELI-ALPS, ELI-HU Non-profit Ltd., Wolfgang Sandner utca 3, H-6728 Szeged, Hungary}%

\author{S\'{a}ndor Varr\'{o}}
\affiliation{ELI-ALPS, ELI-HU Non-profit Ltd., Wolfgang Sandner utca 3, H-6728 Szeged, Hungary}%
\affiliation{Wigner Research Centre for Physics, Konkoly-Thege
	M. \'ut 29-33, H-1121 Budapest, Hungary}

\begin{abstract}
We present a study of  photon statistics associated with high-order harmonic generation (HHG) involving one-mode and intermodal correlations of the high harmonic photons. The aim of the paper is to give insight into the nonclassical properties of high-order harmonic modes. To this end, we use a simplified model describing an elementary quantum emitter: the model of a two-level atom.  While the material system is extremely simplified in this description, the conclusions and the methods may be generalized for more complex cases.
Our primary interest is an effective model of HHG in which the exciting pulse is classical, and the harmonics are quantized, although we touch upon the more generalized, fully quantized model as well.
Evolution of the Mandel-parameter, photon antibunching, squeezing and cross-correlations are calculated.
Results imply that with respect to a single quantized emitter, nonclassicality of the harmonics is present: sub-Poissonian photon statistic and squeezing can characterize certain optical modes, while strong anticorrelation can also be present.
\end{abstract}


\maketitle

\section{Introduction} \label{intro}
\paragraph*{} High-order harmonic generation (HHG) is a strongly nonlinear effect that is observed in several state-of-the-art experiments \cite{F88, IKC05,RSP17}. One of the most important applications is the generation of attosecond pulses, which can monitor or induce physical processes on an experimentally unprecedented time scale \cite{FT92,BCR09}. Therefore, deep understanding of the physical mechanisms underlying the phenomenon of HHG is of crucial importance.
\\ Usually, the calculations involved in strong-field physics and attosecond science are based on the semiclassical approach, treating the electron quantum-mechanically, and the electromagnetic
field classically \cite{K64,KI09,L94}.
\par On the other hand, quantized description of various phenomena in strong fields has already been discussed in the early ’80s. Reference \cite{JV81} gives a non-perturbative treatment of HHG in the nonlinear Compton process using a fully quantized framework \cite{photonics8070269}. More recently, the idea that the photon number distribution of a laser pulse shows fingerprints of the generation of high-order harmonics after the interaction with matter appeared in a theoretical paper \cite{GO16}. Later on, the effect has been demonstrated both with gaseous \cite{TNKIGIT17} and solid-state targets \cite{TKDF19}. A short review of quantum-optical spectrometry is contained in \cite{photonics8060192}. On the theoretical side, a general perturbative treatment of the problem has been given in \cite{AG19}. 

In the most widely used picture that describes gaseous targets, the continuum energy levels and the charge acceleration plays an important role \cite{L94}. 
However, for solid-state targets, it is possible for only  bound states to be populated during the process \cite{G10}, and even the two-level approximation  can  be  valid  for  quantum  wells \cite{H94}. Previous works highlight how a driven  two-level  system  can model the properties observed in HHG spectra  \cite{GK95,GGK97,FR03,GC16,photonics8070263}.
Let  us  note  that a model with finite number of bound states  can  directly  be  related  to the harmonic generation of solid-state targets described in velocity gauge using single-particle and dipole approximations. Then all transitions are interband, that is, the dynamics of states with different $\mathbf{k}$ eigenvalues are independent \cite{MS11,Foldi17}. 

\bigskip

An interesting experimental aspect of HHG is the possibility of  performing photon counting experiments. In order to obtain exact photon statistics, one should calculate all higher-order correlation functions, but the experimentally most significant terms are those of up to second order \cite{MW95}.
\par Of particular interest are the intermodal cross-correlation functions, the calculation of which is generally nontrivial. The properties of the two-mode correlation function are connected with the characterization of the electromagnetic field as a whole, which, as a first approximation, can be done by measuring second-order intermode cross-correlations.

\bigskip

Naturally, quantum-optical properties like photon statistics are inherently unobtainable from a semiclassical approach.  Although there have been numerous studies --both experimental and theoretical-- about the photon statistics of second-, and Nth order harmonics \cite{AJW98,YM91}, fully quantum optical treatments of the HHG are relatively rare.


\bigskip

The quantum properties of the radiation by an isolated, point-like system may be affected for example by the following properties:
The structure of the relevant energy levels of the system and modes, and the transition dipole-moments; the polarization, intensity, and quantum properties (photon statistics) of the excitation; the timescale of the harmonic generation, i.e. whether spontaneous emission plays role in the dynamics.
\\
In this paper, we will only deal with strong, coherent excitation and its interaction with a two-level system, on a timescale that is short compared to the characteristic time of the spontaneous emission.

\bigskip

The paper is organized in the following way:
In Sec.\ref{corrfunc} we give definitions of the correlation-functions and other quantities calculated in this article. Sec.\ref{model} specifies the model \cite{GC16} we investigate. 
In Sec.\ref{dynequs} we present semi-analytic and numerical results connected to the photon statistics of high-order harmonics induced by classical radiation, while the intermodal correlations are treated in Sec.\ref{Intermodecorr}.
As an outlook, we give a brief presentation of results concerning the quantized excitation in Sec.\ref{Initial}. Conclusions are given in Sec.\ref{conclusion}.

\section{Correlation functions}\label{corrfunc}

The complete characterization of the radiation field in terms of intensity is only possible in limited cases. More accurate descriptions are possible by using a hierarchy of correlation functions as  defined  by  optical  coherence  theory \cite{CG69a,CG69ab,MW95}.  

Correlation functions provide a concise method for expressing the degree to which two (or more) dynamical properties are correlated. 
Generally speaking, the response of a system to a specific weak probe is often directly related to a correlation function, therefore the determination of specific correlation functions have been the focus of many experimental settings and theoretical investigations \cite{BH07,BW93}.

\bigskip

In quantum optical experiments, the most relevant auto- and cross-correlation functions are between photon numbers.
\\
Usually, semiconductor avalanche  photodiodes  are  used  as  detectors  in  these  experiments \cite{Hama17}.  These detectors typically can achieve  time resolution of the  order  of  500-50  ps.  
Since detectors typically average  over  the detection time, fast fluctuations of  the  correlation  function (which can contain important information concerning the physics of ultrafast processes) are blurred. In recent years, picosecond resolution has become possible  \cite{MFJ10,BIG15}. 

\bigskip
Below, we introduce quantities relevant to quantum optical experiments. The experimental setup to measure these quantities is typically similar to that of Hanbury Brown and Twiss \cite{HT56}. 

\textbf{Mandel Q parameter}
\begin{equation*}
Q_n(t) \equiv
\langle N_n(t) \rangle \bigg(
g^2_n(t,0)-1
\bigg) 
= \frac {(\Delta N_n)^2}{\langle N_n \rangle}-1
\end{equation*}
Whenever it takes negative values, the photon statistics is called sub-Poissonian and can be called nonclassical \cite{L83}.
We note that during the time-evolution, there can be time instants when the photon number expectation value becomes (exactly or numerically) zero. This circumstance can cause difficulties during numerical evaluations of $Q_n$.

The definition of the Q-parameter is related to the second-order coherence function $g^2_n(t,\tau)$, specifically for the one-time $\tau=0$ case. The second-order coherence function
\begin{equation*}
g^2_i(t,\tau) \equiv
\dfrac{\langle a^\dagger_i(t) N_i(t\!+\!\tau) a_i(t) \rangle}{\langle N_i(t) \rangle \langle N_i(t\!+\!\tau) \rangle }
\sim \dfrac{P(t+\tau|t)}{P(t+\tau)}
\end{equation*}
is related to the conditional probability $P(t+\tau|t)$ of a detector measuring a second photon at time $t+\tau$, granted that a first photon was measured at $t$.

\textbf{Photon antibunching measure}
\begin{equation}\label{antibunchmeasure}
\delta g^2_i(t,\tau) \equiv
\lim\limits_{\tau\to 0}
\dfrac{g^2_i(t,\tau)-g^2_i(t,0)}{\tau} 
\end{equation} 

Definitions and quantifications of photon bunching and antibunching are not completely unambiguous in the literature. Especially in experimental situations, when one considers the integration of signals by the detector, the concept of bunching needs careful handling \cite{PJ91,SS83,ZM90,PH94,GK05,AJW98}.
For the sake of clarity, we list the commonly used definitions of photon antibunching for a single mode.

\begin{figure}[h!]\hspace*{-0.2cm}
	\centering
	\includegraphics[width=1.02\linewidth]{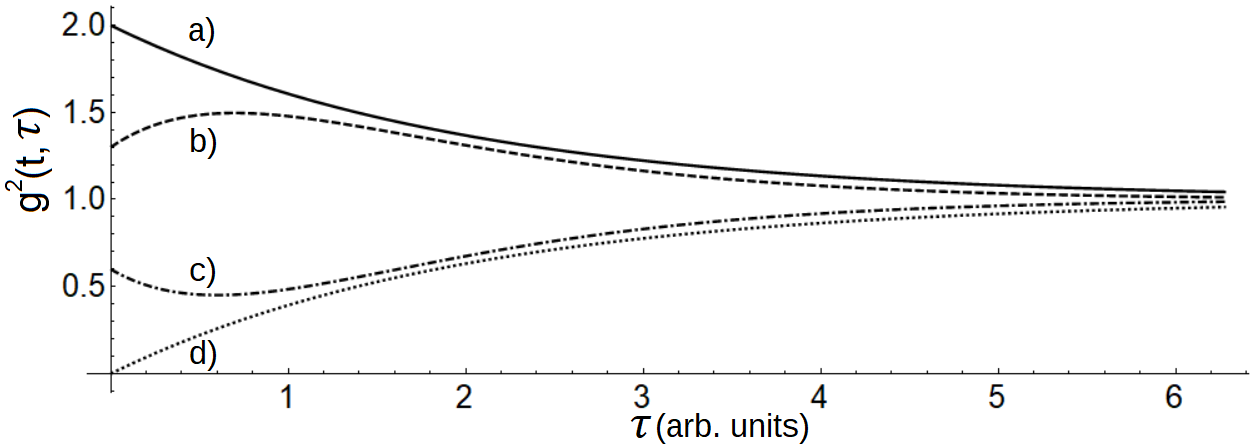}
	\caption[Sketch of two-time correlation functions.]{Two-time correlation functions with fixed $t$. a) super-Poissonian bunching, b) super-Poissonian  antibunching, c) sub-Poissonian bunching,  d) sub-Poissonian antibunching.}
	\label{fig:ketido1}
\end{figure}

The  presence of photon antibunching is equivalent to:
\\-Def.1)  $g^2_n(t,0)\!<\!1$ or $Q(t)\!<\!0$ \cite{MS91}.
\\~~ -Def.2) 
$G^2_n(t, \tau)\!>\!G^2_n(t,0)$, where the quantities are defined by
$G^2_n(t, \tau)\!=\!\langle a^\dagger_n(t)a^\dagger_n(t\!+\!\tau) a_n(t\!+\!\tau)a_n(t) \rangle$  \cite{MW95}. 
\\~~ -Def.3)  $g^2_n(t,\tau)\!>\!g^2_n(t,0)$ \cite{GK05} . 

Comparison of possible definitions are discussed in \cite{AJW98,MB99,MAM99,MMX10}.
We have used antibunching according to Def.3), similarly to \cite{DSB92}. Illustrative cases are presented in Fig.(\ref{fig:ketido1}).
The positivity of  $\partial_{\tau}  g^2(t,\tau)|_{\tau=0}$ implies (assuming that photon absorption happened at time $t$) that the probability of photon absorption is larger some small $\tau$ time later than the simultaneous absorption of two photons.

\textbf{Intermodal cross-correlation for two modes}
\begin{equation*}
g^2_{ij}(t)=
\dfrac{\langle  N_i(t) N_j(t) \rangle}{\langle N_i(t) \rangle \langle N_j(t) \rangle } 
\end{equation*} 
The field is nonclassical, if the inequality $g^2_{ii}(t) g^2_{jj}(t) < \left(g^2_{ij}(t) \right)^2 $ stands \cite{PJSLGX98}. There are additional inequalities \cite{MMX10}, but here we  only consider the following one: Specifically, nonclassical entanglement between two ($i$ and $j$) modes is implied if the 
\begin{equation*}
\langle N_i N_j \rangle<|\langle a_i a^\dagger_j \rangle|^2
\end{equation*}
relation is fulfilled \cite{HMZM06}.

\bigskip

Another quantity of interest is the squeezing of the harmonic modes. 
Light is considered to be squeezed in a given mode if there exists a quadrature-variance smaller than the one associated with vacuum state \cite{WM94,CMW85}.
The minimal variance (and its associated phase) can be calculated through the smaller eigenvalue (and associated eigenvector) of the noise-ellipse matrix.
To quantify it, we use the following notations:
$
\\X_n \!\equiv \tfrac{a^\dagger_n + a_n}{2},
Y_n \!\equiv i\tfrac{ a^\dagger_n - a_n}{2},
X_{2_n}\!\equiv \tfrac{a^{\dagger 2}_n + a^2_n}{2},
Y_{2_n}\!\equiv i\tfrac{ a^{\dagger 2}_n - a^2_n}{2}.
$
Then the noise-ellipse is:
\begin{align}
\begin{pmatrix}
\langle (\Delta X)^2 \rangle & \tfrac{1}{2}\langle\{\Delta X, \Delta Y\}\rangle \\
\tfrac{1}{2}\langle\{\Delta X, \Delta Y\}\rangle & \langle (\Delta Y)^2 \rangle
\end{pmatrix}
\end{align}
where  $\{\cdot~,\cdot\}$ denotes the anticommutator.
The eigenvalues \cite{PJ91,LPP88}, expressed with the above notations are:
\begin{align}\label{variances}
\lambda_{\pm}=
\dfrac{1}{4}\bigg[ 
\langle \{ \Delta a, \Delta a^\dagger \} \rangle
\pm 2 |\langle (\Delta a)^2 \rangle|
\bigg]
\nonumber \\
= \dfrac{1}{4}\bigg[ 
1 \!+ \!2\big(\langle N \rangle \!-\! \langle X \rangle^2 \!-\! \langle Y \rangle^2 \big)
\!\pm 2 | \langle X_2 \! + \! i Y_2 \rangle \!-\! \langle X \!+\! i Y \rangle^2 |
\bigg] .
\end{align}
The quantum state of a given mode is squeezed if $\lambda_-<\tfrac{1}{4}$.

\section{Model}\label{model}
In our investigation, we assumed that the excitations are --at least before interaction-- characterized by coherent states.  
The model of the material is a two-level system. The simplicity of two-level systems helps to form qualitatively (and sometimes quantitatively) correct predictions, and offer insight into the dynamics of the HHG. Furthermore, the methods used in this article can be  generalized  to  more  complex  high  harmonic  sources as well.

Although harmonic generation is a nonlinear optical effect, only relevant in high-field settings, the intensities of the harmonics are typically much lower than that of the excitation. Therefore, especially when investigating the "one-atom response", the assumption of classicality for the scattered harmonic radiation might not be necessarily valid. 

\bigskip

Let us consider the following Hamiltonian terms:
\begin{align*}
H_{a}=\hbar\dfrac{\omega_{0}}{2}\sigma_z,
\\
H_{h}=\sum_{n\in HH} \hbar \omega_n a_n^{\dagger}a_n, ~~
H_{ah}=\sum_{n\in HH} \hbar \frac{\Omega_n}{2}\sigma_x (a_n+a_n^{\dagger}),
\\
H_{e}=\sum_{n\in E} \hbar \omega_n a_n^{\dagger}a_n, ~~
H_{ae}=\sum_{n\in E} \hbar \frac{\Omega_n}{2}\sigma_x (a_n+a_n^{\dagger}),
\label{Hint}
\end{align*}
where the first term corresponds to a two-level atom, with the operators $\sigma_i$ being the Pauli-matrices, the second(fourth) term describes  the quantized scattered(excitation) electromagnetic modes, and the third(fifth) term expresses the interaction using dipole approximation. Summations over HH(E) refers to summation over modes of the high harmonics(excitation).

The general Hamiltonian, with both quantized harmonics and excitation can be written as: 
\begin{equation}\label{HQQ}
H_{qq}=H_{a}+H_{ah}+H_{h}+H_{ae}+H_{e}.
\end{equation}

Usually, it is assumed that the interaction of the pulse with matter changes the quantum statistics of the pulse only slightly. In our experience the backaction on the excitation, while can be noticeable, does not (at least for short interaction times) have significant effect on the dynamics of the dipole operator.
For this reason, we will be utilizing the classical approximation of the excitation, leading to the following effective Hamiltonian \cite{GC16}:
\begin{equation}
H_{qc}(t)=H_{a}+H_{h}+H_{ah}+H_{ex}(t)
\label{Ham}.
\end{equation}
Where the electromagnetic field of the excitation can be described as: 
\begin{equation}
H_{ex}(t)= -DE(t)=-d\sigma_x E(t)=\hbar\frac{\Omega(t)}{2} \sigma_x.
\end{equation}
We note that $\Omega_n\!=\!2d\sqrt{\frac{\hbar \omega_n}{\epsilon_0 V}}$, where $V$ is the quantization volume.
Let us denote the eigenstates of the atomic Hamiltonian by $|e\rangle$ and $|g\rangle$, 
i.e., $H_a|e\rangle=\tfrac{\hbar\omega_0}{2} |e\rangle,$ $H_a|g\rangle\!=-\tfrac{\hbar\omega_0}{2} |g\rangle$.

\bigskip

In this model, in principle all the electromagnetic modes would need to be accounted for, with proper initial conditions. For simplicity, in numerical calculations we assumed the initial condition to be $|\Psi(0)\rangle \!=\!|g\rangle \!\otimes\! |0\rangle \!\otimes\dots \otimes \!|0\rangle$.

The dynamics of the high-order harmonics are induced by the quantized dipole driven by strong classical and weak quantized fields. For practical reasons, we need to utilize some kind of approximation for the calculations.
\par While the semiclassical theory of radiation assumes  that the atomic quantities and field quantities are independent \cite{AL87} i.e. $\langle \sigma_i a_j\rangle=\langle \sigma_i \rangle \langle a_j\rangle$, that assumption is clearly unacceptable when one investigates correlation-functions. 
\\
On the other hand, it is true that the effect of the low-intensity high-harmonic radiation on a classically driven dipole --and thus on each other-- is weak. 
The cumulative effects of the mode-mode interactions become palpable at the timescale of the spontaneous emission lifetime. Since in experimental settings the pulse is in the order of femtoseconds, we will neglect the interplay between different harmonics and consider the high harmonic modes independently.

We will treat the cases of monochromatic and pulsed excitations separately, both assumed to be linearly polarized. The pulsed excitations have an electric field:
\begin{align*}
E(t)=A\sin^2(\omega_e t)\sin(\omega_f t+\phi) ~~~\text{if}~~~ t\in[0,\pi/\omega_e]
\\ ~~~ 0 ~~~~~~~~~~  \text{otherwise} ~~,
\end{align*} 
where $A$ denotes the amplitude, $\phi$ is the carrier envelope-phase, and $\omega_e\!\ll\!\omega_f$ is fulfilled.

\bigskip

In this article, we use the following notations.
\\Photon number operator: $N_{n}$;\\
atomic operators:
$U=\sigma_{x}, 
V=-\sigma_{y} , 
W=\sigma_{z};$
\\first and second order field-operators:
\\$(a^\dagger_n + a_n),
 i(a^\dagger_n - a_n), 
(a^{\dagger2}_n + a_n^2), 
 i(a^{\dagger2}_{n} - a_n^2);$
\\first-order atom-field operators:
$
\\U_{n}^{\pm}=(i)^{(1\mp1)/2}\sigma_{x}(a_{n} \pm a_{n}^{\dagger}), \\V_{n}^{\pm}=-(i)^{(1\mp1)/2}\sigma_{y}(a_{n} \pm a_{n}^{\dagger}) , \\W_{n}^{\pm}=(i)^{(1\mp1)/2}\sigma_{z}(a_{n} \pm a_{n}^{\dagger}).
$

\section{One-mode properties}\label{dynequs}
The harmonic spectrum is composed of odd-order harmonics, and Hyper-Raman lines (Mollow-sidebands) which, in this paper, we call even-order harmonics in accordance with our earlier work \cite{GC16}. To clarify our nomenclature, it is worth pointing out that these spectral lines correspond to Mollow-triplets around the odd harmonics \cite{MB70}. Since our focus is on strong-field excitations, the non-odd harmonic optical lines will be near to the spectral position of the even multiples of the base harmonics, so for the sake of simplicity, we will call these (dual) lines even-harmonics. We introduce the notation $\delta\omega$ for the spectral distance between the optical lines and the closest even-order multiple of the base harmonic. The actual value of delta omega is determinded by the parameters of the excitation (amplitude, detuning). Further details can be found in Appendix [\ref{Analy}].

\bigskip

It can be argued \cite{MWENO05} that harmonics should be defined based on the phase and carrier-frequency, (through $\chi^k$ nonlinear susceptibility) without reference to the position in the optical spectrum. From this argument, the radiation which we call even-harmonics should be more precisely called odd-harmonics disguised as even harmonics \cite{TTMOW03}.

We note that according to \cite{2019APS..DMPS01062J,JTJBA20,XYJA16}, results obtained for molecular targets suggests that the intensity of Mollow sidebands remain around the same order of magnitude as odd-harmonics (in the macroscopic spectrum). They are also radiated at wider angles, therefore can be distinguished from the main harmonics, and in principle can also be isolated. 
It is important to note that recent experiments observed these hyper-Raman lines for atomic targets as well \cite{BEBDPLMM19}.

\bigskip

Starting from this section, we will present numerical results first, followed by the analytically gained ones, the mathematical background of which can be found in the Appendix. 

\subsection{Mandel-parameter}

\begin{figure}[h!]
	\centering
	\includegraphics[width=1.0\linewidth]{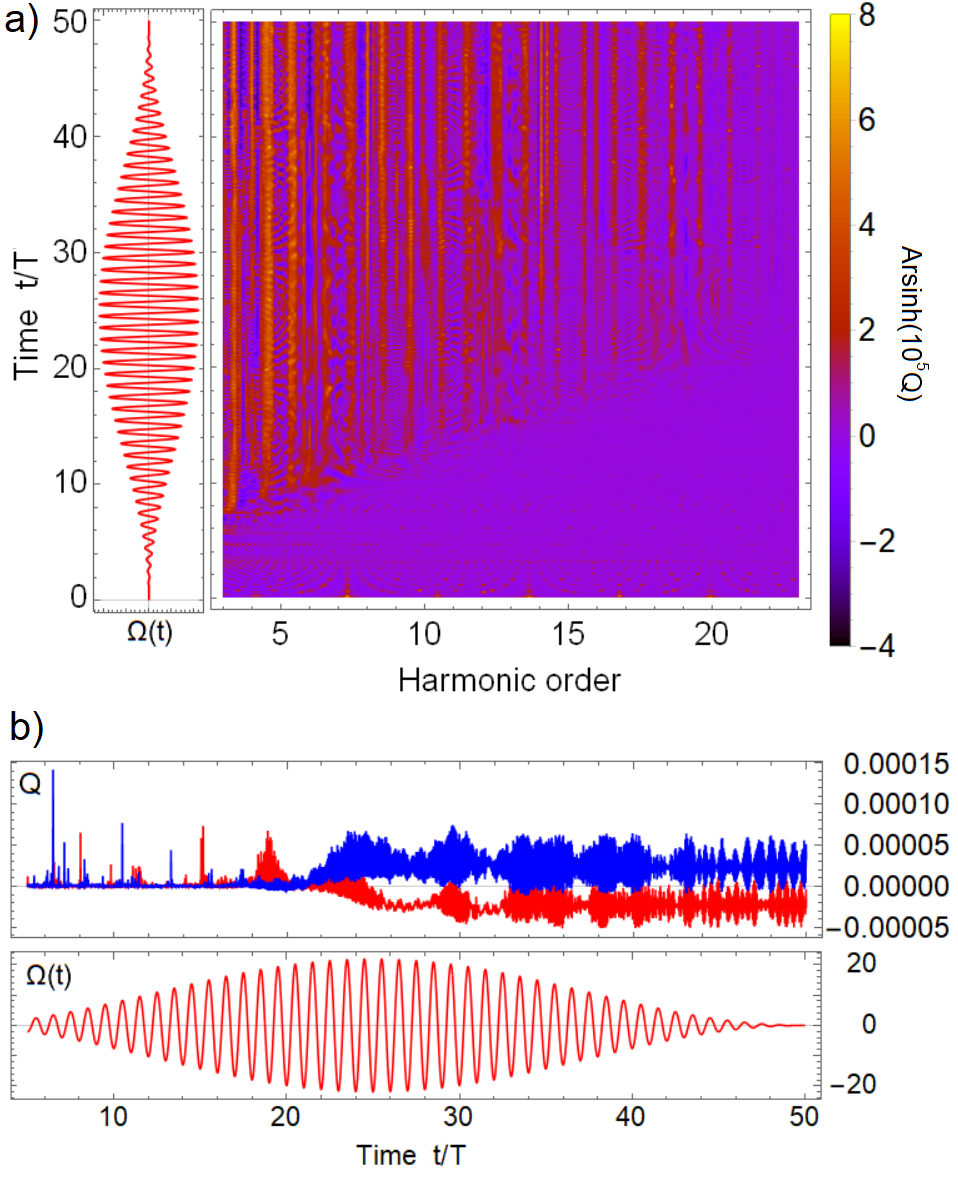}
	\caption[Mandel-parameter under pulsed excitation.]{Mandel-parameter under pulsed excitation. Subfigure \textbf{a)} shows $\sinh^{-1}(10^5Q)$ on the central panel, with the (resonant) excitation on the left panel.  Top panel of subfigure \textbf{b)} shows time-evolution of $Q$ of an even (18th) harmonic with the $\omega_0/\omega$ detuning being 0.8 (blue) and 1.2 (red), while the bottom panel shows the excitation.}
	\label{fig:mandel2}
\end{figure}

When we consider pulsed excitation, the photon statistic --just like the spectrum-- becomes hard to characterize.
The narrow peaks of the Mandel-parameters at the early stage of the time evolution, visible in  Fig.(\ref{fig:mandel2}/b) are signatures of the initial transient effects. Then rapid oscillations appear, which have considerably more regular pattern --that is, the average over multiple cycles is constant-- when the pulse is over ($t>\tau$). 
Note that the detuning has strong effect on the photon statistical properties in the case of pulsed excitations.

To make the analysis transparent, we present results specific to monochromatic excitations. 
We note that in Fig.(\ref{fig:mandel}), the visible super-Poissonian modes (positioned between the harmonics) are characterized by very low photon-number expectation values, and their super-Poissonian quality can be considered a numerical artefact.

\bigskip

The Mandel-parameters of odd harmonic modes usually displays both positive and negative values within an optical cycle, being approximately zero on average.
Even harmonic lines have more complicated behaviour: 

\bigskip

-If $\delta\omega\!=\!0$, the single even harmonic line will first become sub-Poissonian, then super-Poissonian [Fig.(\ref{fig:mandel}/c)]. An intuitive explanation is that this optical line develops significant squeezing [See (\ref{quadraturedyn})] while the photon-number mean value is only increasing moderately, rendering the photon-number fluctuation large as the interaction time increases.

\bigskip

-If $|\delta\omega|$ is large, usually only one of the even-harmonic modes will be significantly populated, while the other spectral line develops significantly sub-Poissonian statistics [Fig.(\ref{fig:mandel}/d)]. Our calculation shows that $Q\approx-\langle N\rangle$, which is equivalent to $\langle N^2 \rangle\approx \langle N \rangle$, in other words, the photon statistic of such even harmonic mode is essentially a superposition of zero and one-photon states.
\begin{widetext}
	
	\begin{minipage}{\linewidth}
		\begin{figure}[H]
	\centering
	\includegraphics[width=0.88\linewidth]{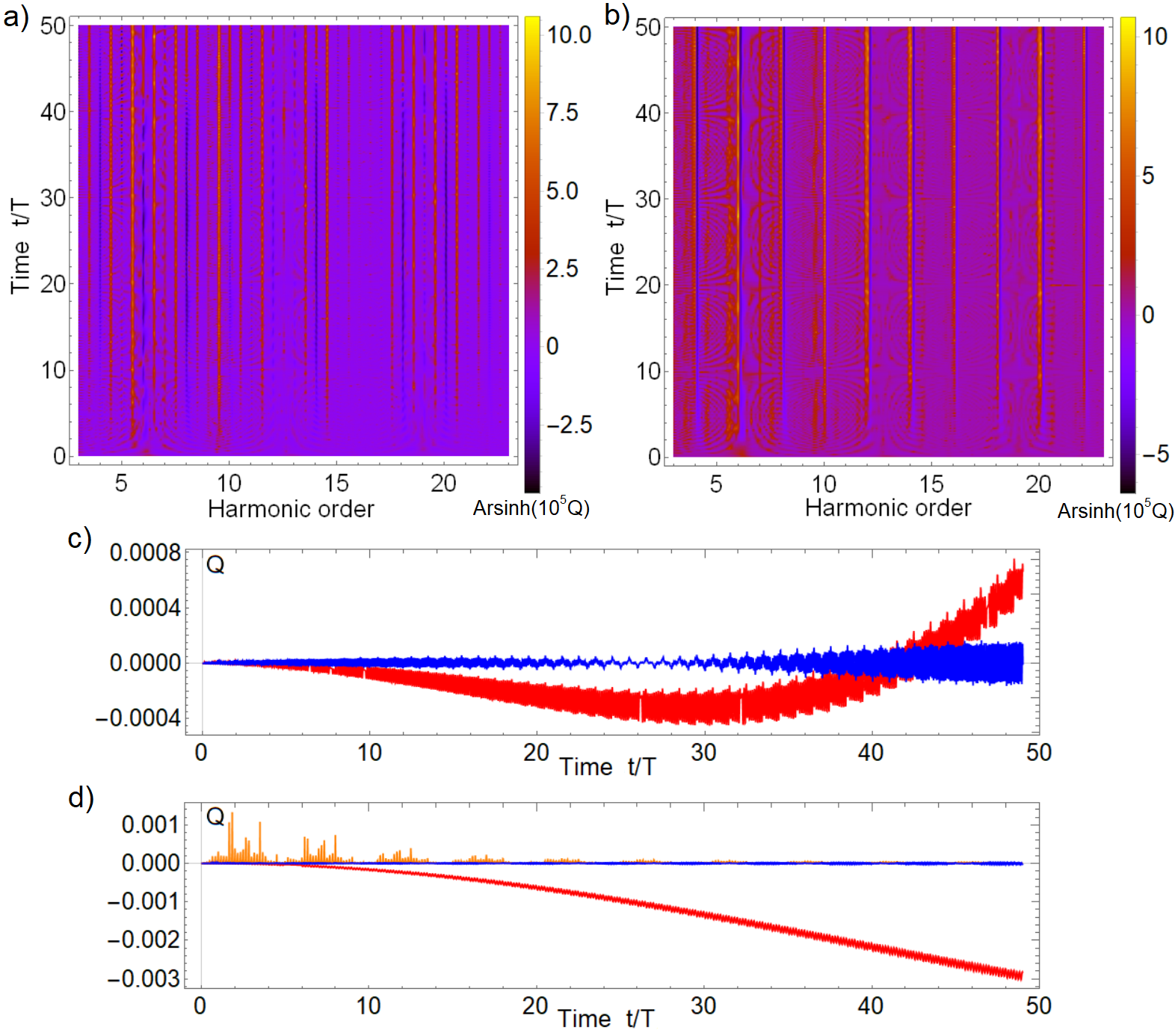}
	\caption{Mandel-parameter under monochromatic excitations. The parameters are choosen such that $\delta\omega=0$ on subfigure $\textbf{a)}$ and $\textbf{c)}$ and $\delta\omega=0.1\omega$ on subfigure $\textbf{b)}$ and $\textbf{d)}$. Subfigure $\textbf{a)}$ and $\textbf{b)}$ shows evolution of the rescaled Mandel-parameter $\sinh^{-1}(10^5Q)$. 
	Subfigure $\textbf{c)}$ and $\textbf{d)}$  shows time-evolution of the Mandel-parameter of given odd-harmonic line (blue) and even harmonic line(s) (red and orange). The mode corresponding to the orange line has negligible population.}
	\label{fig:mandel}
		\end{figure}    
\end{minipage}

\end{widetext}

These results imply that sub-Poissonian behaviour is present in the even-harmonic (Hyper-Raman) lines. In order to check that our results remain valid for the case of higher photon numbers, we considered a very long, nearly monochromatic pulsed excitation.

\bigskip

Note that our earlier results in \cite{GC16} were considerably less detailed. There, we only considered a single mode, while the excitation was pulsed, which did not offer as much transparency as the analysis of monochromatic excitation does.

\bigskip

We give the asymptotic value of Mandel-parameters as a function of harmonic order, for a long "boxed" monochromatic excitation in Fig.(\ref{fig:longpulsemandel}). The excitation is 500 optical cycle long, with additional 25 cycle duration of rise and decay. In order to filter out the oscillations after the excitation, we took the average value over many cycles. 

With respect to these averaged asymptotical values, we found our statements to be, up to a good approximation, valid. It is worth noting however that in these calculations the odd-harmonics themselves could display super-, or sub-Poissonian behaviour, but for them, the $|Q|\ll\langle N\rangle$ relation is fulfilled. 

\begin{widetext}
	
	\begin{minipage}{\linewidth}
	\begin{figure}[H]
	\centering
	\includegraphics[width=0.9\linewidth]{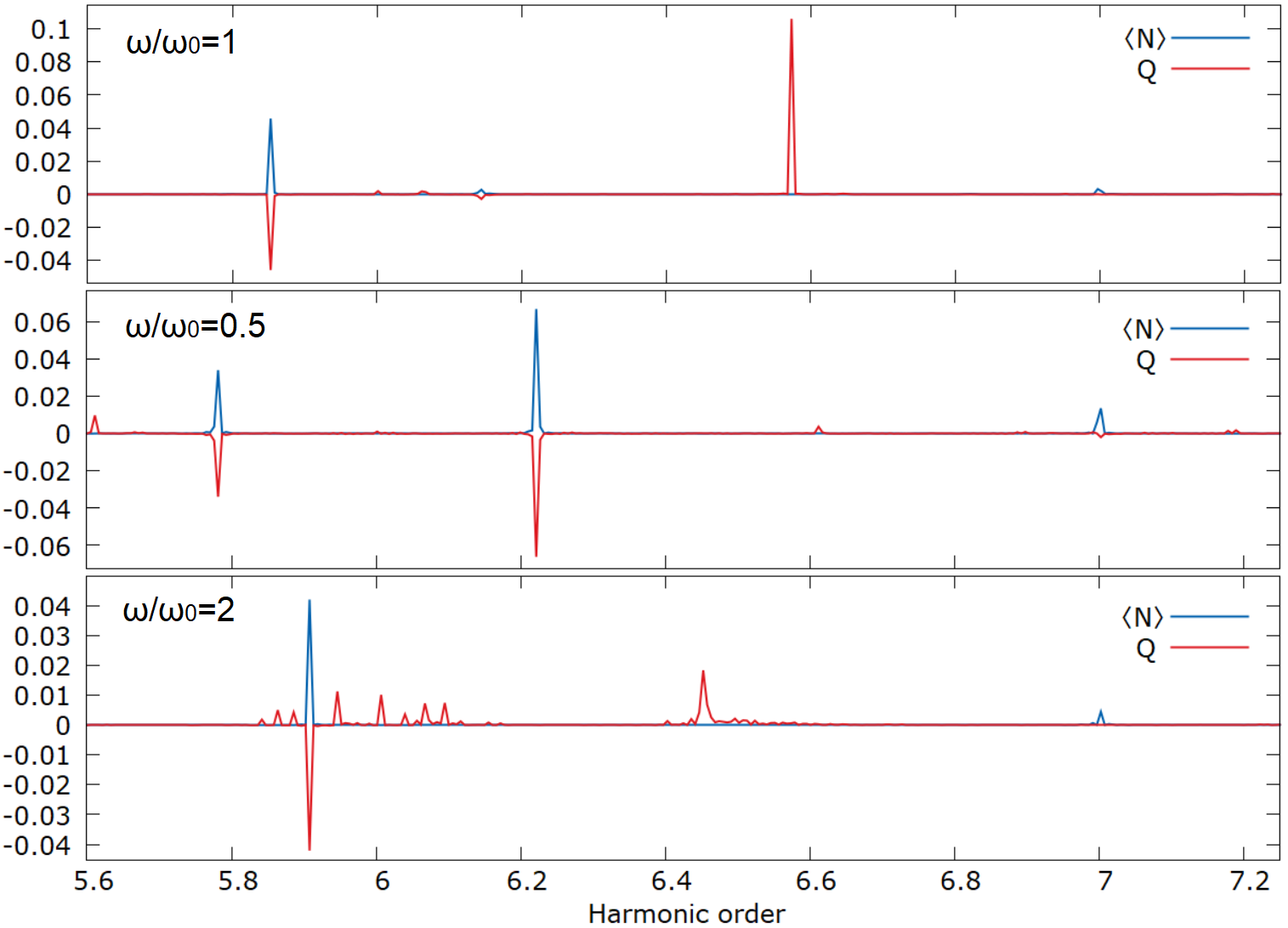}
	\caption{Asymptotical averaged photon number expectation values (blue) and Mandel-parameters (red), for resonant, red-detuned, and blue-detuned excitations. The horizontal axis is shared among the subfigures. The excitation is pulsed, with 25 optical cycle duration of rise and decay, and 500 cycle duration of constant amplitude.}
	\label{fig:longpulsemandel}
		\end{figure}    
\end{minipage}
\end{widetext}

As we can observe, the presence of even-order harmonic (hyper-Raman line) modes with significant population (and nonclassicality) holds also for such pulsed excitations. As long as the rise and decay of the amplitude is relatively short, the sensitivity to the detuning is practically eliminated.
\subsection{Time-evolution of quadrature-variance}\label{quadraturedyn}
The quadrature variance-spectrum, that is, the value of the minimal and maximal variances $\lambda_{\pm}$ in Eq.(\ref{variances}) as the function of the parameter $\omega_n$, can display distinct properties depending on the chosen  parameters, but it can be summarized in the following way:

The odd-harmonics on the plateau display squeezing, which, however, is usually weak. The even harmonics have quantum properties that are very sensitive to the excitation parameters.
If the parameters are chosen such that $\delta\omega$ is practically zero, particularly strong squeezing is present, among the even harmonics, whereas if $\delta\omega$ is large, strongly anti-squeezed states ($\lambda_->\tfrac{1}{2}$) will be produced, mainly in the more populated modes of the even harmonic lines. 
\\
For illustration, we plotted relevant quantities in Fig.\ref{fig:squeeze}.

\bigskip

One may conclude, that as far as producing squeezed states are concerned (obviously within the limits of this model) special attention is to be given to the parameter space for which $\delta\omega=0$ (for details, see the Appendix). In such cases, the even harmonic lines display unusual behaviour compared to other lines. The squeeze in these modes will grow faster than the photon number expectation values, rendering the photon statistics super-Poissonian beyond a given interaction time [see Fig.(\ref{fig:mandel}/c)].

\begin{widetext}
	
	\begin{minipage}{\linewidth}
		\begin{figure}[H]
			\centering
			\includegraphics[width=1.0\linewidth]{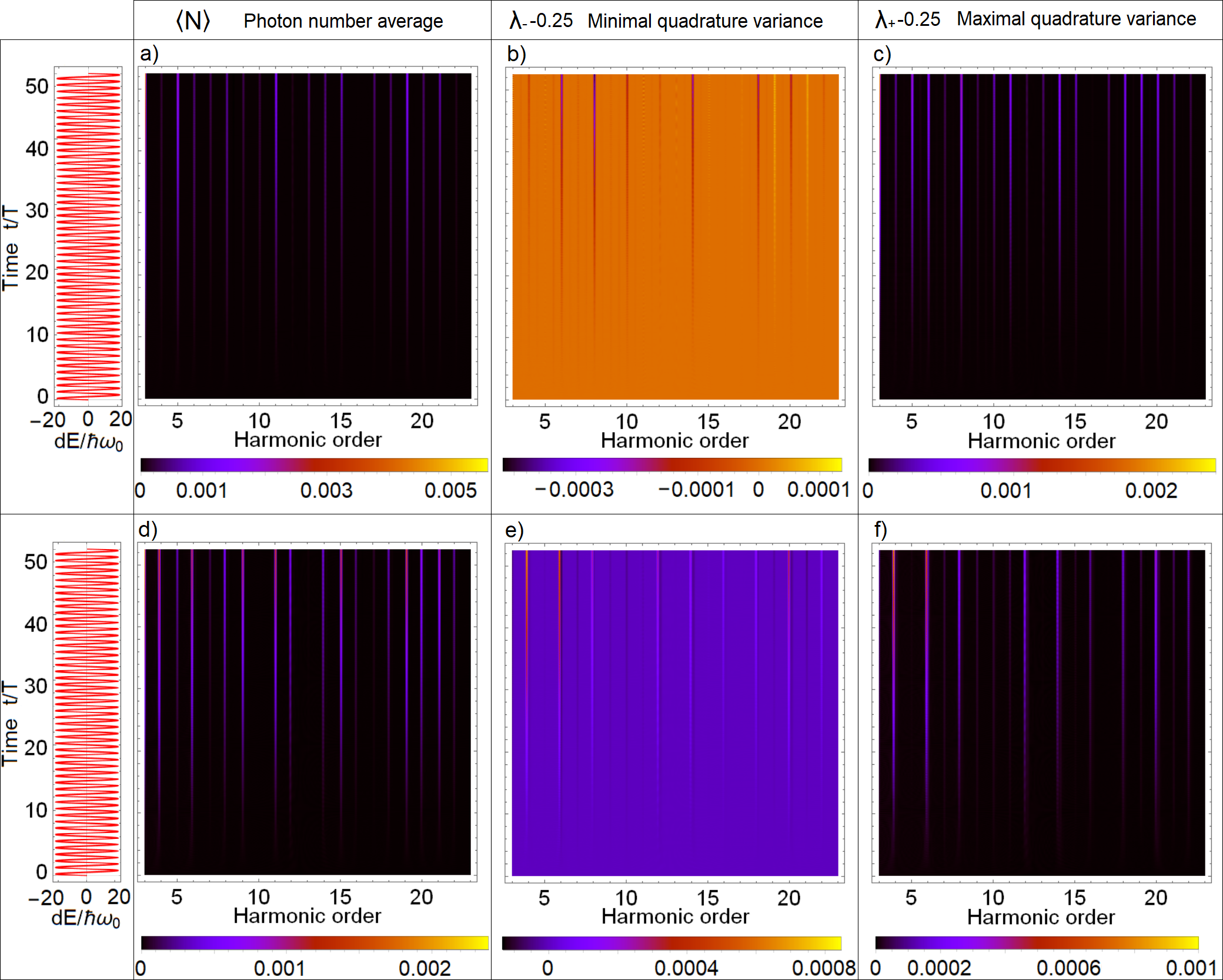}
			\caption{Time evolution of field-quantities. The upper and lower row of figures corresponds to two monochromatic excitations with slightly different amplitudes  (shown in the left panels). Time evolution of $\delta\lambda_{-}-\tfrac{1}{4}$ is in subfigure \textbf{b)} and \textbf{e)}, while $\delta\lambda_{+}-\tfrac{1}{4}$ can be observed in subfigure \textbf{c)} and \textbf{f)}. For the sake of clarity, we showed $\langle N\rangle$ in subfigures \textbf{a)} and \textbf{d)}. Vertical axis is time measured in $T$, horizontal axis is harmonic order. Amplitudes of the excitation has been choosen so that $\delta\omega=0$ for \textbf{a)} to \textbf{c)} and $\delta\omega=0.1\omega$ for  \textbf{d)}, \textbf{e)} and \textbf{f)}, corresponding to dimensionless amplitudes $21.2119$ and $21.8354$ respectively.}
			\label{fig:squeeze}
		\end{figure}    
	\end{minipage}
	
\end{widetext}

\subsection{Analytical approximation of photon statistics for extremal $\delta\omega$ values}\label{Analyticalkorrelacio}
Limiting ourselves to the case of monochromatic excitation, it is possible to give analytic insight into the resulting photon statistics of individual harmonics. Specifically, we focus on the special set of parameters that grant extremal $\delta\omega$.

\bigskip

Perturbative calculation in the fourth order, applied to the electromagnetic mode with $\omega_n$ angular frequency, leads us to the expression below. The $b^e_{2}$ and $b^g_{2}$ coefficients correspond to the (two-photon) components ($|e\rangle|2\rangle$ and $|g\rangle|2\rangle$) of the quantum state.
\\The $\mathcal{F}$ symbol denotes Fourier-transform, while $\zeta_{1/2}$ are terms given in the Appendix.
\begin{widetext}
\begin{eqnarray}
b^e_{2}(t)^{(4)} \approx
-\dfrac{\Omega^2_n t^2}{4\sqrt{2}} 
\bigg(  ~\mathcal{F}\!\left[ b^e_{0}(0)[1 + i\zeta_1] -  ib^g_{0}(0) \zeta_2 \right]\!
\big(\!-\omega_n + \delta\omega \big)  
~\mathcal{F}\big[1-i\sigma_1 \big] (-\omega_n -\delta\omega) 
\nonumber\\ 
- e^{-i2\phi_0}  ~\mathcal{F}\!\left[ b^g_{0}(0)[1 - i\zeta_1] -  ib^e_{0}(0) \zeta^*_2 \right]\!
\big(\!-\omega_n - \delta\omega \big)
~\mathcal{F}\big[i\zeta^*_2 \big](-\omega_n -\delta\omega) \bigg) \hspace{0.5cm}
\\
b^g_{2}(t)^{(4)} \approx
-\dfrac{\Omega^2_n t^2}{4\sqrt{2}} 
\bigg( ~\mathcal{F}\!\left[ b^g_{0}(0)[1 - i\zeta_1] -  ib^e_{0}(0) \zeta^*_2 \right]\!
\big(\!-\omega_n - \delta\omega \big) 
~\mathcal{F}\big[1+i\zeta_1\big](-\omega_n + \delta \omega)
\nonumber\\
-e^{i2\phi_0}  ~\mathcal{F}\!\left[ b^e_{0}(0)[1 + i\zeta_1] -  ib^g_{0}(0) \zeta_2 \right]\!
\big(\!-\omega_n + \delta\omega \big) 
~\mathcal{F}\big[ i\zeta_2 \big](-\omega_n + \delta \omega) \bigg). \hspace{0.5cm}
\end{eqnarray}
\end{widetext}
The perturbative approach leads us to the the following statements:

Odd-harmonics: The quantum state of the field, up to the order presented can be written as $|\Psi\rangle^{(4)}_{HH}\approx |0\rangle + \alpha|1\rangle + \alpha^2/\sqrt{2}|2\rangle$, where $\alpha\!\in\!\mathbb{C}$.
By assuming that the coefficients follow a similar pattern at higher-order, the quantum state of odd-order harmonics would be a coherent state with label $\alpha=\Omega_n/2 \mathcal{F}[\zeta_2](-\omega_n \pm\delta\omega) t$. Naturally, the perturbative calculation is not strictly true. Predicted intensities are smaller than the numerically calculated values, but the nearly-Poissonian (on the average) statistics is correctly reproduced.

Even-harmonics: The quantum state of both Hyper-Raman lines can be written as $|\Psi\rangle^{(4)}_{HH}\approx |0\rangle + \beta|1\rangle$ where, if we choose initial condition $|\Psi\rangle_a(0)=|g\rangle$, the $\beta$ parameter is nonzero only for one even harmonic line, characterized by sub-Poissonian statistics. The  $Q=-\langle N\rangle$ relation observed in numerical calculations follows straightforwardly, since $\langle N^2\rangle = \langle N\rangle$.

\subsection{Photon-bunching properties}

By the definition that we use, photon (anti)bunching is implied by the sign of $\partial_{\tau}  g^2(t,\tau)$.
Expanding  $g^2(t,t+\tau)-g^2(t,t)$ up to the first order in $\tau$, we get:
\begin{align}\label{bunching1order}
g^2(\tau)-g^2(0) 
\approx
\tau \dfrac{\Omega }{2} \dfrac{\langle a^\dagger U^- a\rangle \langle N \rangle  
	-	\langle a^\dagger N a \rangle \langle  U^- \rangle }
{\langle N \rangle \langle N \rangle \langle N +\tau \tfrac{\Omega }{2} U^- \rangle },
\end{align} 
which can be simplfied further.
Since $\langle N \rangle >0$ and $\langle a^\dagger N a  \rangle >0$ in the limit of small $\tau$, the presence of photon bunching is determined by the sign and relative magnitude of $\langle U^- \rangle \! \propto\! \langle \dot{N} \rangle$ and $\langle a^\dagger U^- a\rangle$.
\begin{figure}[h!]
	\centering
	\includegraphics[width=1.0\linewidth]{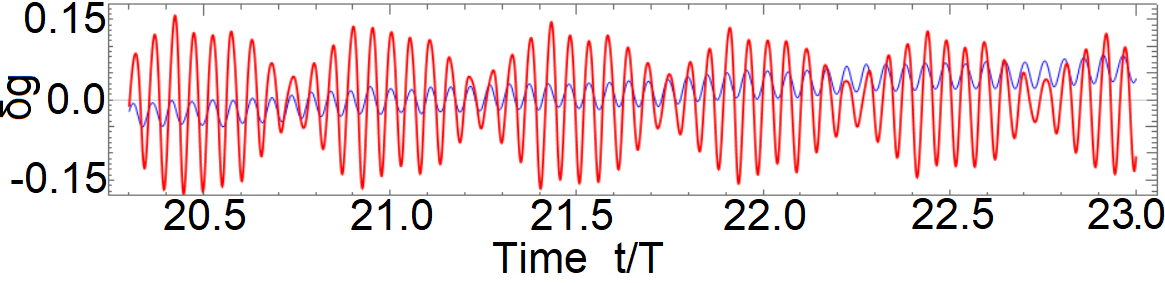}
	\caption[Time-evolution of photon-antibunching measure of an even harmonic.]{Time-evolution of photon-antibunching measure [Eq.(\ref{antibunchmeasure})] of a given even harmonic mode (red), with the monochromatic excitation parameters chosen so that $|\delta\omega|$ is large. For the purpose of illustration we also show the rescaled and displaced mean photon value in blue.}
	\label{fig:photonbunching}
\end{figure}

\bigskip

The only significant dynamics that we observed  --both for pulsed and monochromatic excitation-- is the roughly $2\pi/\omega_n$ periodic photon antibunching. For most modes, the average value of $\delta g$ is indistinguishable from zero, but for the sub-Poissonian even harmonic that have been plotted on Fig.(\ref{fig:photonbunching}), the average is slightly negative. Considering quantities averaged over optical cycles, the fact that the sub-Poissonian statistic coincides with photon bunching, can be interpreted as a consequence of the non-stationary state of the mode as the Mandel-parameter decreases in time.

\section{Intermode correlation between harmonics}\label{Intermodecorr}
\subsection{Numerical results}

As a first step of characterizing the radiation field as a whole, i.e., calculating the emerging cross-correlations, the two-mode approximation proves to be useful.
\\
First, we treat the case of monochromatic excitation, with parameters choosen so that $|\delta\omega|=0$ and $|\delta\omega|=0.1\omega$ is fulfilled. Results can be seen on Fig.(\ref{fig:keresztkorr1}) and Fig.(\ref{fig:keresztkorr2}) respectively.

\bigskip

Specifically for the $|\delta\omega|=0$ parameter --which we can associate with particularly significant squeezing being present within the even harmonic modes-- numerical results imply that:
\\-Odd harmonic photons tend to be correlated with other odd harmonics, generally close to the classical limit [i.e. correlation of unit value];
\\-Even and odd harmonic photons tend to be anti-correlated;
\\-Even harmonic photons tend to be strongly correlated.
This implies the theoretical possibility that HHG can be the source of wideband, correlated squeezed states. 

\begin{widetext}
	
	\begin{minipage}{\linewidth}
\begin{figure}[H]
	\centering
	\includegraphics[width=0.98\linewidth]{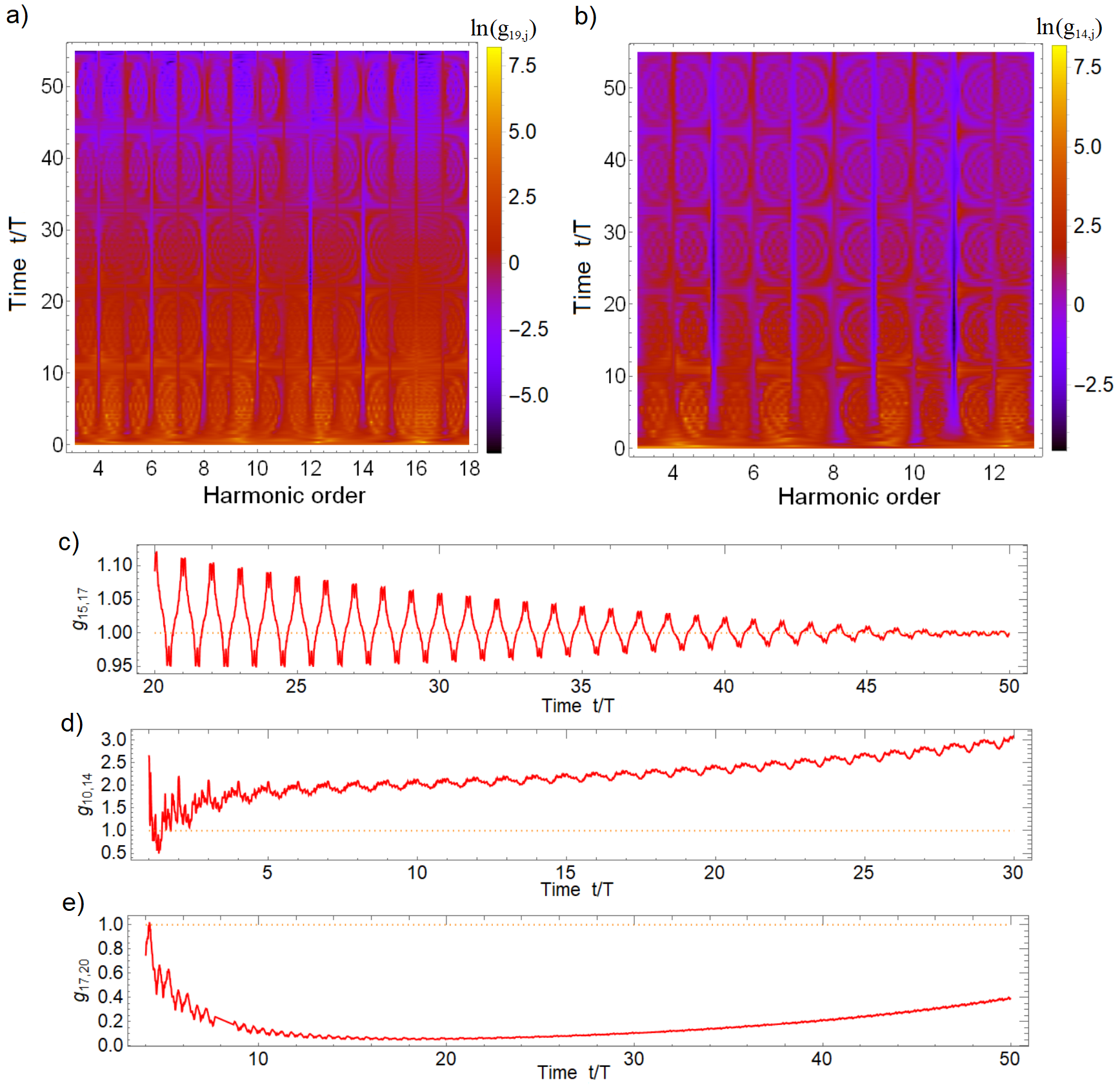}
	\caption[Plot of cross-correlation functions for monochromatic excitation]{Plot of cross-correlation functions for monochromatic excitation, with $\delta\omega=0$. $\mathbf{a)}$ shows the time and frequency dependence of the two-mode correlation on logarithmic scale, where one of the modes is the 19th harmonic. $\mathbf{b)}$ Same as in subfigure a), but the fixed mode is the 14th harmonic. $\mathbf{c)}$ Cross-correlation between two odd (15th and 17th) harmonics.  $\mathbf{d)}$ Cross-correlation between two even (10th and 14th) harmonics. $\mathbf{e)}$ Cross-correlation between odd and even (17th and 20th) harmonics.}
	\label{fig:keresztkorr1}
\end{figure}
	\end{minipage}

\end{widetext}

\bigskip

For $|\delta\omega|=0.1\omega$ parameter --which is relatively close to the maximal $|\delta\omega|$ case, which we can associate with particularly nonclassical, nearly one-photon states being present within the even harmonic modes-- numerical results imply that:
\\-Odd harmonic photons tend to be correlated with other odd harmonics, generally close to the classical limit;
\\-Even and odd harmonic photons tend to be significantly anti-correlated;
\\-Even harmonic photons tend to be strongly anti-correlated.
This implies the theoretical possibility that HHG can be the source of one-photon states in wide spectral ranges. We note that (at least in the cases investigated by us) the modes are more populated (the quantum states are closer to one-photon states) if the interaction time is longer, and the coupling is stronger.  

\begin{widetext}
	
	\begin{minipage}{\linewidth}
		\begin{figure}[H]
	\centering
	\includegraphics[width=0.98\linewidth]{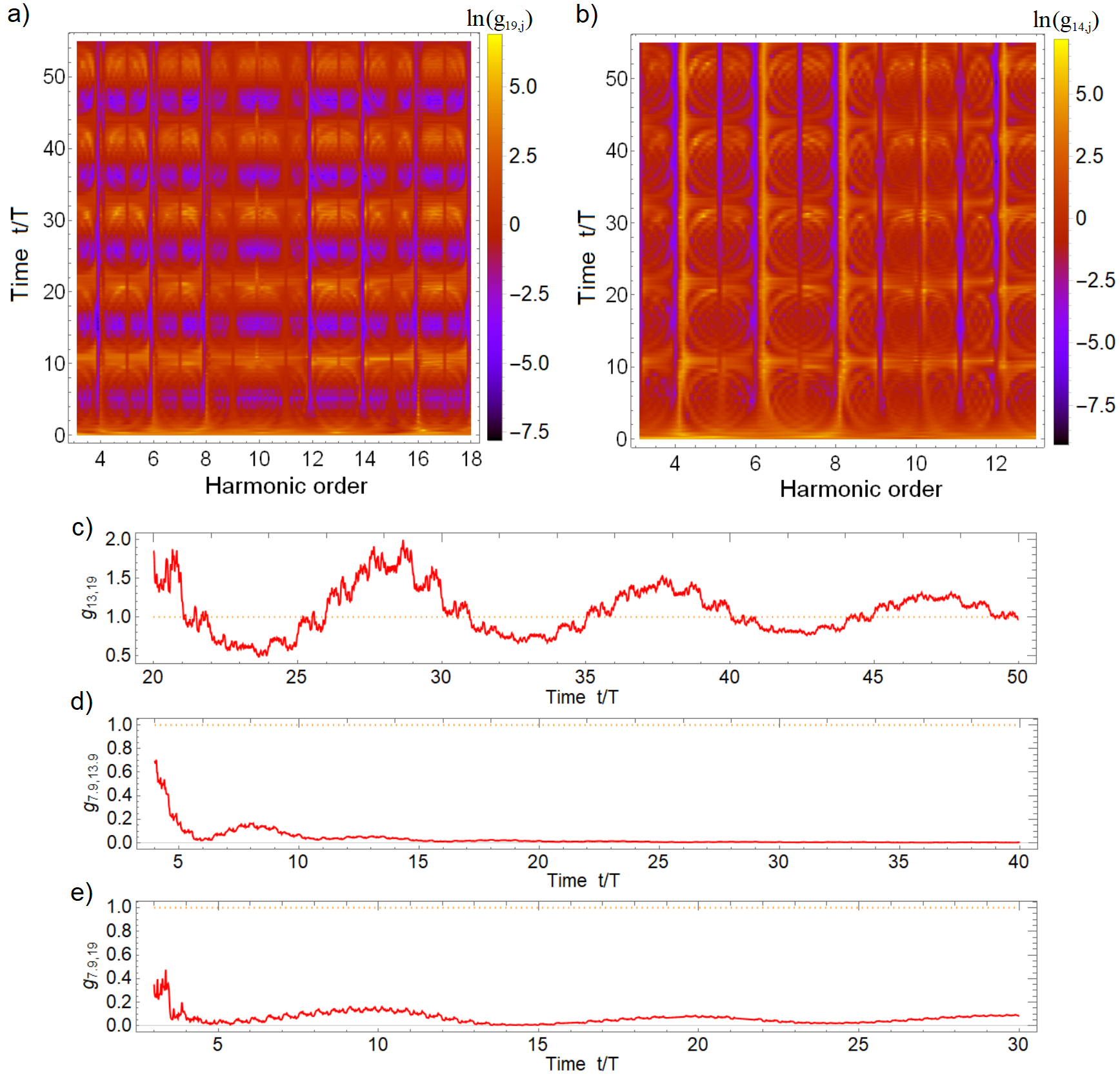}
	\caption[Plot of cross-correlation functions for monochromatic excitation]{Plot of cross-correlation functions for monochromatic excitation, with $\delta\omega=0.1\omega$. $\mathbf{a)}$ shows the time and frequency dependence of the two-mode correlation on logarithmic scale, where one of the mode is the 19th harmonic. $\mathbf{b)}$ Same as in subfigure a), but the fixed mode is that of the 14th harmonic's more populated line. $\mathbf{c)}$ Cross-correlation between two odd (13th and 19th) harmonic.  $\mathbf{d)}$ Cross-correlation between two more populated even (8th and 14th) harmonic lines. $\mathbf{e)}$ Cross-correlation between odd and even (8th and 19th) harmonics.}
	\label{fig:keresztkorr2}
\end{figure}
\end{minipage}

\end{widetext}

For pulsed excitation, the cross-correlations become more complicated [see Fig.(\ref{fig:cross})], but certain qualitative statements can be made:
\\-Odd harmonic photons tend to be correlated with other odd harmonics, generally close to the classical limit;
\\-Even and odd harmonic photons tend to be significantly anti-correlated;
\\-Even harmonics photon's correlation with other even harmonic photons can be either stronger or weaker than the classical limit.

\begin{widetext}
	
	\begin{minipage}{\linewidth}
		\begin{figure}[H]
	\centering
	\includegraphics[width=0.99\linewidth]{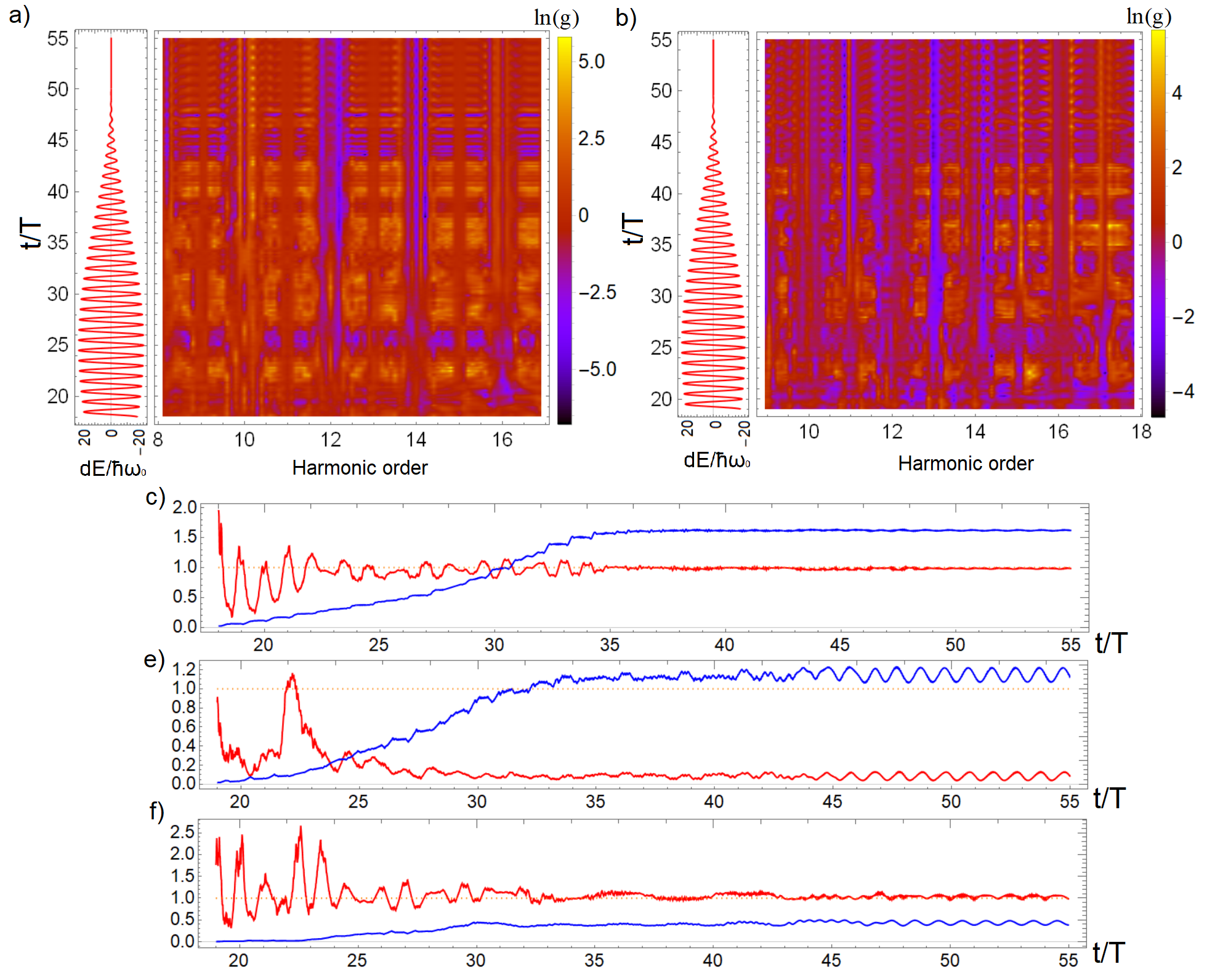}
	\caption[Plot of cross-correlation functions for pulsed excitation.]{Plot of cross-correlation functions for pulsed excitation. The left side shows the time dependence of the excitation in units of the optical cycles. $\mathbf{a)}$ Logarithm of the cross-correlation between the 17th mode and other optical modes.  $\mathbf{b)}$ Logarithm of the cross-correlation between the  18th harmonic mode and other optical modes.  $\mathbf{c)-e)}$ Cross-correlation function (red) and sum of ($10^4\times$)photon number expectation values (blue) for: $15th$ and $17th$ harmonics; $17th$ and $18th$ harmonics; and between two optical lines of the 18th harmonic; respectively. }
	\label{fig:cross}
\end{figure}
\end{minipage}

\end{widetext}

\subsection{Analytical approximation of intermodal correlations for extremal $\delta\omega$ values}\label{Multimode}

An approximate physical picture of the quantum state of the scattered radiation in the case of monochromatic excitation can be constructed using analytic methods. As in the previous section, we will only consider the special cases of extremal $\delta\omega$.

Since our goal here is to gain a simple physical picture  we will assume that the state of each electromagnetic mode spans a minimal space containing $|0\rangle$ and $|1\rangle$ photon number states, that is, we focus only on the terms with dominant contributions in intermodal cross-correlations.
For the sake of transparency, we will consider two electromagnetic modes (of arbitrary $\omega_1$ and $\omega_2$ respectively). 
The quantum state can be written explicitely as:
\begin{align}\hspace{-0.6cm}
|\Psi\rangle =
U e^{-\tfrac{i}{\hbar}\epsilon_+ t} \bigg(
b^{e}_{00}(t)  | \tilde{e} \rangle |0\rangle |0\rangle
+
b^{e}_{01}(t)  | \tilde{e} \rangle |0\rangle |1\rangle e^{-i\omega_2 t}
\nonumber \\ +
b^{e}_{10}(t)  | \tilde{e} \rangle |1\rangle |0\rangle e^{-i\omega_1 t}
+
b^{e}_{11}(t)  | \tilde{e} \rangle |1\rangle |1\rangle e^{-i(\omega_1 +\omega_2 )t}
\bigg)
\nonumber \\
+
U e^{-\tfrac{i}{\hbar}\epsilon_- t} \bigg(
b^{g}_{00}(t) |\tilde{g} \rangle |0\rangle |0\rangle
+
b^{g}_{01}(t) |\tilde{g} \rangle |0\rangle |1\rangle e^{-i\omega_2 t}
\nonumber \\ +
b^{g}_{10}(t) |\tilde{g} \rangle |1\rangle |0\rangle e^{-i\omega_1 t}
+
b^{g}_{11}(t) |\tilde{g} \rangle |1\rangle |1\rangle e^{-i(\omega_1+\omega_2) t} \bigg) ,
\end{align}
where $U\equiv e^{i\tfrac{A\xi}{2\omega}\sin(\omega t + \phi_0)\sigma_x}  e^{\tfrac{i}{2}(\omega t + \phi_0)\sigma_z}$ [for details, see the Appendix]. We fix the initial condition so that at time $t=0$,  $b^g_{lm}\!=\!b^e_{lm}\!=\!0$ unless  $l=m=0$.
The time-evolution of the $b$ coefficients is induced by

\begin{align}
\hbar W(t) 
+ \hbar\left(\sigma^+ e^{i(\omega t + \phi_0)} 
+ \sigma^- e^{-i(\omega t + \phi_0)} \right) 
\nonumber\\
\times\bigg[ \dfrac{\Omega_1}{2}(a^\dagger_1 + a_1)
+ \dfrac{\Omega_2}{2}(a^\dagger_2 + a_2) \bigg],
\end{align}
where $W(t)$ is defined by (\ref{Wdef}).
Below we give the perturbative results, neglecting second-order contributions of $\hbar W(t)$, and all nonresonant terms.

In first-, and second-order, we get:
\begin{widetext}
\begin{eqnarray}
b^e_{00}(t)^{(1)} = b^e_{00}(0)[1 + i\zeta_1(t)] -  ib^g_{00}(0) \zeta_2(t) \hspace{3cm}
\\
b^g_{00}(t)^{(1)} = b^g_{00}(0)[1 - i\zeta_1(t)] -  ib^e_{00}(0) \zeta^*_2(t) \hspace{3cm}
\\
b^e_{10}(t)^{(2)} =
-i\dfrac{\Omega_1 t}{2}e^{-i\phi_0}  ~\mathcal{F}\!\left[ b^g_{00}(0)[1 - i\zeta_1] -  ib^e_{00}(0) \zeta^*_2 \right]\!
\big(\!-\omega_1 - \delta\omega \big)   \hspace{1.6cm}
\\
b^g_{10}(t)^{(2)} = -i\dfrac{\Omega_1 t}{2}e^{i\phi_0}  ~\mathcal{F}\!\left[ b^e_{00}(0)
[1 + i\zeta_1] -  ib^g_{00}(0) \zeta_2 \right]\!
\big(\!-\omega_1 + \delta\omega \big)   \hspace{1.7cm}
\\
b^e_{01}(t)^{(2)} =
-i\dfrac{\Omega_2 t}{2}e^{-i\phi_0}  ~\mathcal{F}\!\left[ b^g_{00}(0)
[1- i\zeta_1] -  ib^e_{00}(0) \zeta^*_2 \right]\!
\big(\!-\omega_2 - \delta\omega \big)   \hspace{1.6cm}
\\
b^g_{01}(t)^{(2)} = 
-i\dfrac{\Omega_2 t}{2}e^{i\phi_0}  ~\mathcal{F}\!\left[ b^e_{00}(0)
[1 + i\zeta_1] -  ib^g_{00}(0) \zeta_2 \right]\!
\big(\!-\omega_2 + \delta\omega \big)   \hspace{1.7cm}
\end{eqnarray}
The most important third-order effect is that induced by $\hbar W(t)$, between the populations $b^g_{10}$,$b^e_{10}$ and $b^g_{01}$,$b^e_{01}$, resulting in:
\begin{eqnarray}
b^e_{10}(t)^{(3)} \approx
-i\dfrac{\Omega_1 t}{2} \bigg( e^{-i\phi_0}  ~\mathcal{F}\!\left[ b^g_{00}(0)[1 - i\zeta_1] -  ib^e_{00}(0) \zeta^*_2 \right]\!
\big(\!-\omega_1 - \delta\omega \big) ~[1+i\zeta_1]
\nonumber\\
-e^{i\phi_0}  ~\mathcal{F}\!\left[ b^e_{00}(0)[1 + i\zeta_1] -  ib^g_{00}(0) \zeta_2 \right]\!
\big(\!-\omega_1 + \delta\omega \big) ~[i\zeta_2] \bigg)
\\
b^g_{10}(t)^{(3)} \approx
-i\dfrac{\Omega_1 t}{2} \bigg( e^{i\phi_0}  ~\mathcal{F}\!\left[ b^e_{00}(0)[1 + i\zeta_1] -  ib^g_{00}(0) \zeta_2 \right]\!
\big(\!-\omega_1 + \delta\omega \big)  ~[1-i\zeta_1]
\nonumber\\
- e^{-i\phi_0}  ~\mathcal{F}\!\left[ b^g_{00}(0)[1 - i\zeta_1] -  ib^e_{00}(0) \zeta^*_2 \right]\!
\big(\!-\omega_1 - \delta\omega \big)
~[i\zeta^*_2] \bigg)
\end{eqnarray}
\begin{eqnarray}
b^e_{01}(t)^{(3)} \approx
-i\dfrac{\Omega_2 t}{2} \bigg( e^{-i\phi_0}  ~\mathcal{F}\!\left[ b^g_{00}(0)[1 - i\zeta_1] -  ib^e_{00}(0) \zeta^*_2 \right]\!
\big(\!-\omega_2 - \delta\omega \big) ~[1+i\zeta_1]
\nonumber\\
-e^{i\phi_0}  ~\mathcal{F}\!\left[ b^e_{00}(0)[1 + i\zeta_1] -  ib^g_{00}(0) \zeta_2 \right]\!
\big(\!-\omega_2 + \delta\omega \big) ~[i\zeta_2] \bigg)
\\
b^g_{01}(t)^{(3)} \approx
-i\dfrac{\Omega_2 t}{2} \bigg( e^{i\phi_0}  ~\mathcal{F}\!\left[ b^e_{00}(0)[1 + i\zeta_1] -  ib^g_{00}(0) \zeta_2 \right]\!
\big(\!-\omega_2 + \delta\omega \big)  ~[1-i\zeta_1]
\nonumber\\
- e^{-i\phi_0}  ~\mathcal{F}\!\left[ b^g_{00}(0)[1 - i\zeta_1] -  ib^e_{00}(0) \zeta^*_2 \right]\!
\big(\!-\omega_2 - \delta\omega \big)
~[i\zeta^*_2] \bigg)
\end{eqnarray}
Then, at the fourth order we get:
\begin{eqnarray}
b^g_{11}(t)^{(4)} \approx
-\dfrac{\Omega_1 \Omega_2 t^2}{8} \bigg( \mathcal{F}\!\big[ b^g_{00}(0)[1 - i\zeta_1] -  ib^e_{00}(0) \zeta^*_2 \big]\! 
\big(\!-\omega_1 - \delta\omega \big) 
~\mathcal{F}\!\big[1+i\zeta_1 \big] \big(\!-\omega_2 + \delta\omega \big) 
\nonumber\\
- e^{i2\phi_0} \mathcal{F}\!\big[ b^e_{00}(0)[1 + i\zeta_1] -  ib^g_{00}(0) \zeta_2 \big]\!
\big(\!-\omega_1 + \delta\omega \big) 
~\mathcal{F}\!\big[i\zeta_2\big] \big(\!-\omega_2 + \delta\omega \big) 
\nonumber\\
+ ~\mathcal{F}\!\big[ b^g_{00}(0)[1 - i\zeta_1] -  ib^e_{00}(0) \zeta^*_2 \big]\!
\big(\!-\omega_2 - \delta\omega \big) 
~\mathcal{F} [1+i\zeta_1] \big(\!-\omega_1 + \delta\omega \big) 
\nonumber\\
-e^{i2\phi_0}  ~\mathcal{F}\big[ b^e_{00}(0)[1 + i\zeta_1] -  ib^g_{00}(0) \zeta_2 \big]\!
\big(\!-\omega_2 + \delta\omega \big)
~\mathcal{F} [i\zeta_2]  \big(\!-\omega_1 + \delta\omega \big) 
\bigg),
\hspace*{0.6cm}
\end{eqnarray}
\begin{eqnarray}
b^e_{11}(t)^{(4)} \approx
-\dfrac{\Omega_1 \Omega_2 t^2}{8} 
\bigg( \mathcal{F}\!\big[ b^e_{00}(0)[1 + i\zeta_1] -  ib^g_{00}(0) \zeta_2 \big]\! 
\big(\!-\omega_1 + \delta\omega \big) 
~\mathcal{F}\!\big[1-i\zeta_1 \big] \big(\!-\omega_2 - \delta\omega \big) 
\nonumber\\
- e^{-i2\phi_0} \mathcal{F}\!\big[ b^g_{00}(0)[1 - i\zeta_1] -  ib^e_{00}(0) \zeta^*_2 \big]\!
\big(\!-\omega_1 - \delta\omega \big) 
~\mathcal{F}\!\big[i\zeta^*_2\big] \big(\!-\omega_2 - \delta\omega \big) 
\nonumber\\
+ ~\mathcal{F}\!\big[ b^e_{00}(0)[1 + i\zeta_1] -  ib^g_{00}(0) \zeta_2 \big]\!
\big(\!-\omega_2 + \delta\omega \big) 
~\mathcal{F} [1-i\zeta_1] \big(\!-\omega_1 - \delta\omega \big) 
\nonumber\\
-e^{-i2\phi_0}  ~\mathcal{F}\big[ b^g_{00}(0)[1 - i\zeta_1] - ib^e_{00}(0) \zeta^*_2 \big]\!
\big(\!-\omega_2 - \delta\omega \big)
~\mathcal{F} [i\zeta^*_2]  \big(\!-\omega_1 - \delta\omega \big) 
\bigg).
\hspace*{0.6cm}
\end{eqnarray}
\end{widetext}
Due to the coefficients constituting a quickly decreasing series, the intermode photon cross-correlation between two ($\omega_1,\omega_2$) high-order harmonic mode can be reasonably represented by:
\begin{equation*}
g_{12}(t)\approx
\dfrac{|b^e_{11}|^2 + |b^g_{11}|^2}
{(|b^e_{10}|^2 + |b^g_{10}|^2)(|b^e_{01}|^2 + |b^g_{01}|^2)},
\end{equation*}
where, during evaluation, it is worth separating the special cases below. For the sake of simplicity, we will consider only the initial condition $|g\rangle|0\rangle\dots|0\rangle$, which does not limit the validity of the conclusions.

\textbf{Odd-odd harmonic modes:}
Let the frequencies be $\omega_1= (2k_1+1) \omega$ and $\omega_2= (2k_2+1) \omega$, where $k_1,k_2\in\mathcal{N}$. Between two odd harmonic lines, the cross-correlation is 
\begin{widetext}
\begin{align}\hspace{-1cm}
g_{12}(t)\approx 
\dfrac{ \frac{\Omega_1^2\Omega_2^2 t^4}{64}
	\big| 2 e^{-i2\phi_0}
	\mathcal{F}[\zeta^*_2](\!-\omega_1\!-\!\delta\omega)
	\mathcal{F}[\zeta^*_2](\!-\omega_2\!-\!\delta\omega)
	\big|^2 }
{\frac{\Omega^2_1 t^2}{4}
	\big|e^{-i\phi_0} \mathcal{F}[\zeta^*_2](\!-\omega_1\!-\!\delta\omega) \big|^2
	\frac{\Omega^2_2 t^2}{4}
	\big|e^{-i\phi_0} \mathcal{F}[\zeta^*_2](\!-\omega_2\!-\!\delta\omega) \big|^2 }
= 1 ,
\end{align}
\end{widetext}
which is close to the numerically calculated value.

\textbf{Even-even harmonic modes:} 
Let us choose the frequencies as $\omega_1= 2k_1\omega+\delta\omega$ and $\omega_2= 2k_2\omega+\delta\omega$. Between such even harmonic lines, the cross-correlation is 
\begin{widetext}
\begin{align}\hspace{-1cm}
g_{12}(t)\approx 
\dfrac{ \frac{\Omega_1^2\Omega_2^2 t^4}{64}
	\big|
	\mathcal{F}[\zeta_1](\!-\omega_1\!+\!\delta\omega)
	\mathcal{F}[\zeta_1](\!-\omega_2\!-\!\delta\omega)
	+
	\mathcal{F}[\zeta_1](\!-\omega_1\!-\!\delta\omega)
	\mathcal{F}[\zeta_1](\!-\omega_2\!+\!\delta\omega)
	\big|^2 }
{\frac{\Omega^2_1 t^2}{4}
	\big|e^{i\phi_0} \mathcal{F}[\zeta_1](\!-\omega_1\!+\!\delta\omega) \big|^2
	\frac{\Omega^2_2 t^2}{4}
	\big|e^{i\phi_0} \mathcal{F}[\zeta_1](\!-\omega_2\!+\!\delta\omega) \big|^2 }
= 0.
\end{align}
\end{widetext}
We can check that the perturbative calculation predict a nonclassical entanglement between even harmonic modes, since $\langle N_1 N_2 \rangle = 0 < |b^{g}_{10}b^{g}_{01}|^2 = |\langle a_1 a^\dagger_2 \rangle|$.

\textbf{Odd-even harmonic modes:}
To calculate the correlation between odd and even harmonics, let us choose the mode-frequencies as $\omega_1= (2k_1+1) \omega$ and $\omega_2= 2k_2\omega+\delta\omega$.
\begin{widetext}
\begin{align}\hspace{-1cm}
g_{12}(t)\approx 
\dfrac{ \frac{\Omega_1^2\Omega_2^2 t^4}{64}
	\big| 
	\mathcal{F}[\zeta^*_2](\!-\omega_1\!-\!\delta\omega)
	\mathcal{F}[\zeta_1](\!-\omega_2\!+\!\delta\omega)
	+ e^{i2\phi_0}
	\mathcal{F}[\zeta_2](\!-\omega_1\!+\!\delta\omega)
	\mathcal{F}[\zeta_1](\!-\omega_2\!+\!\delta\omega)
	\big|^2 }
{\frac{\Omega^2_1 t^2}{4}
	\big|e^{i\phi_0} \mathcal{F}[\zeta_1](\!-\omega_2\!+\!\delta\omega) \big|^2
	\frac{\Omega^2_2 t^2}{4}
	\big|e^{-i\phi_0} \mathcal{F}[\zeta^*_2](\!-\omega_1\!-\!\delta\omega) \big|^2 }
=0,
\end{align}
\end{widetext}
and as above, there is nonclassical entanglement between even and odd harmonic modes.

\subsection{Quantum state of the scattered field}
We stress that the above approximate results can only be considered valid for monochromatic excitation, and for a given set of parameters, for which $|\delta\omega|$ is maximal. (That is, for the special case that can be considered optimal for the creation of one-photon states.) 
By collecting the analytically gained results to reconstruct the quantum-state of the scattered electromagnetic field, (within the given approximations) we can write:
\begin{align*}
|\Psi\rangle_{HH} \sim
c_{o}|\alpha_{3\omega} \rangle 
|\alpha_{5\omega} \rangle 
\dots
|\alpha_{(2k+1)\omega} \rangle 
\\+ 
c^{1}_{e}| \beta_{2\omega +\delta\omega} \rangle
+ 
c^{2}_{e}| \beta_{4\omega +\delta\omega} \rangle
\dots +
c^{k}_{e}| \beta_{2k\omega +\delta\omega} \rangle.
\end{align*}
Here, we only denoted those modes that are in a significantly different state than the vacuum, and $|\alpha\rangle$ is a coherent state, while $|\beta\rangle$ denotes a superposition of $|0\rangle$ and $|1\rangle$. 

\bigskip

We note that the quantum state of the odd-harmonics, and even the anticorrelations between odd and even harmonic photons can be generalized for all monochromatic excitations, and even for pulsed excitations (at least to those that we investigated). However, the quantum state of even harmonic modes and the correlations between them strongly depends on the excitation parameters.

\section{On quantized excitation}\label{Initial}
In this section, the quantized nature of the excitation is incorporated into calculational schemes. Instead of containing the time-dependent semiclassical $\Omega(t)$ term, the excitation in the Hamiltonian (\ref{HQQ}) is represented as a set of quantized modes, with coherent states as the initial condition. For this system, we apply the transformation
\begin{align}\label{qqtoqc}
D_{\text{Exc}}\equiv 
\prod_{n\in\text{Exc}} D_n(\alpha_n e^{-i\omega_n t})
\\
|\Psi \rangle' = D^\dagger_{\text{Exc}} |\Psi \rangle
\nonumber \\
H'_{qq} = D^\dagger_{\text{Exc}} H_{qq} D_{\text{Exc}} 
+ i\hbar D_{\text{Exc}} \partial_t D^\dagger_{\text{Exc}}.
\end{align}
The value of $\alpha_n$ in the above transformation is determined by the spectral composition of the excitation. After simplifications, the Hamiltonian can be reduced to:
\begin{align}\label{Hqqtransform}
H_{qq}' =
\hbar\frac{\omega_0}{2} \sigma_z
+ 
\sum_{n\in\text{HH}} \hbar \bigg(\omega_n  a^\dagger_na_n 
+ \dfrac{\Omega_n}{2}
\sigma_x \big( a^\dagger_{n} + a_{n} \big) \bigg) 
\nonumber \\
+
\sum_{n\in\text{E}} \hbar \bigg(\omega_n  A^\dagger_n A_n 
+ \dfrac{\Omega_n}{2}
\sigma_x \big( A^\dagger_{n} + A_{n} \big) \bigg) 
- \dfrac{\Omega(t)}{2} \sigma_x .
\end{align}
Naturally, the driving term $\Omega(t)$ that dominates the time-evolution of the harmonics, is unaffected by the quantum state of the excitation. At the same time, this term drives the base harmonic mode(s) --that is, the modes corresponding to the excitation-- as well, and (considering that the interaction is resonant) the backaction on the excitation can be significant. 

\bigskip

\subsection{Backaction of HHG on quantized excitation}

During the interaction, the photon statistical properties of the excitation modes are dynamically changing. While this (in our experience) has minor effect on the high harmonic spectrum, the modifications taking place in the quantum state of the excitation can be nevertheless experimentally relevant.

We calculated backaction on a single monochromatic excitation. Our results show that the modification is comparatively small if the parameters are chosen in such a way that $|\delta\omega|$ is extremal.
For this perceived behaviour, we give an approximate analytic explanation, by neglecting the high harmonic modes.

Without the harmonics, the Hamiltonian reduces to
\begin{align}
H_{qq}' = 
\hbar \frac{\omega_0}{2} \sigma_z
+ 
\hbar\bigg( \omega  a^\dagger a
+ \dfrac{\Omega}{2}
\sigma_x \big( a^\dagger + a \big)
\nonumber \\
+ \dfrac{\Omega}{2}
\sigma_x 
\underbrace{\big( \alpha^*e^{i\omega t} + \alpha e^{-i\omega t} \big)}_{\frac{A}{2\Omega}\big(e^{i\omega t+\phi_0} + e^{-i\omega t-\phi_0}\big)}  \bigg).
\end{align}
\begin{widetext}
By writing the quantum state as:
\begin{align}
|\Psi\rangle =
e^{i\tfrac{A\xi}{2\omega} \sin(\omega t + \phi_0 )\sigma_x} e^{\tfrac{i}{2}(\omega t+\phi_0)\sigma_z} e^{-\tfrac{i}{\hbar}t\epsilon_+}
\sum^{\infty}_{j=0} b^e_j | \tilde{e} \rangle|j , \alpha e^{-i\omega t}\rangle 
e^{-ij \omega_n t}
\nonumber \\+
e^{i\tfrac{A\xi}{2\omega} \sin(\omega t + \phi_0)\sigma_x}
e^{\tfrac{i}{2}(\omega t+\phi_0)\sigma_z} e^{-\tfrac{i}{\hbar}t\epsilon_-}
\sum^{\infty}_{j=0} b^g_j | \tilde{g} \rangle|j, \alpha e^{-i\omega t} \rangle 
e^{-ij \omega_n t},
\end{align}
The dynamical equations turn out to be:
\begin{align*}
i \dot{b}^e_j(t) = \langle \tilde{e} | W(t) | \tilde{e} \rangle b^e_j(t)
+ \langle \tilde{e} | W(t) | \tilde{g} \rangle 
e^{i\tfrac{\epsilon_+ - \epsilon_-}{\hbar}t} ~ b^g_j(t) 
+ \dfrac{\Omega}{2} 
e^{i(\delta\omega t - \phi_0) } 
\sum_k \langle j|  a + a^\dagger | k\rangle e^{-i\omega(k-j)t}  ~b^g_k(t)
\\- \Omega \cos\theta 
e^{i\frac{\epsilon_+-\epsilon_-}{\hbar}t } \cos(\omega t+\phi_0)
\sum_k \langle j|  a + a^\dagger | k\rangle e^{-i\omega(k-j)t}  ~b^g_k(t)
+ \dfrac{\Omega}{2} \sin(2\theta) \cos(\omega t+\phi_0)
\sum_k \langle j|  a + a^\dagger | k\rangle e^{-i\omega(k-j)t}  ~b^e_k(t)
,
\\
i \dot{b}^g_j(t) = \langle \tilde{g} | W(t) | \tilde{g} \rangle b^g_j(t)
+ \langle \tilde{g} | W(t) | \tilde{e} \rangle e^{-i\tfrac{\epsilon_+ - \epsilon_-}{\hbar}t} ~b^e_j(t) 
+ \dfrac{\Omega}{2} 
e^{-i(\delta\omega t + \phi_0)} 
\sum_k \langle j|  a + a^\dagger | k\rangle e^{-i\omega(k-j)t}  ~b^e_k(t) 
\\- \Omega \cos\theta 
e^{i\frac{\epsilon_--\epsilon_+}{\hbar}t } \cos(\omega t+\phi_0)
\sum_k \langle j|  a + a^\dagger | k\rangle e^{-i\omega(k-j)t}  ~b^e_k(t)
- \dfrac{\Omega}{2} \sin(2\theta) \cos(\omega t+\phi_0)
\sum_k \langle j|  a + a^\dagger | k\rangle e^{-i\omega(k-j)t}  ~b^g_k(t).
\end{align*}
\end{widetext}
The notations $\theta$ and $\epsilon_\pm$ are defined in the Appendix.
It is easy to check, that unlike in the case of harmonic modes, the dynamical equations regarding the excitation mode have a resonant term (proportional to $\sin(2\theta)$ above) already at first-order perturbation.

\bigskip

To quantify the backaction, --that is, the difference from the initial coherent quantum state that develops over time-- let us use the weighted sum $B_A\equiv \sum_j j\big(|b^e_j|^2+|b^g_j|^2\big)$. Since the (displaced) vacuum-state corresponds to the initially coherent state, $B_A$ characterizes measures the components orthogonal to the coherent state.
\par
\begin{figure}[h!]
	\centering
	\includegraphics[width=1.0\linewidth]{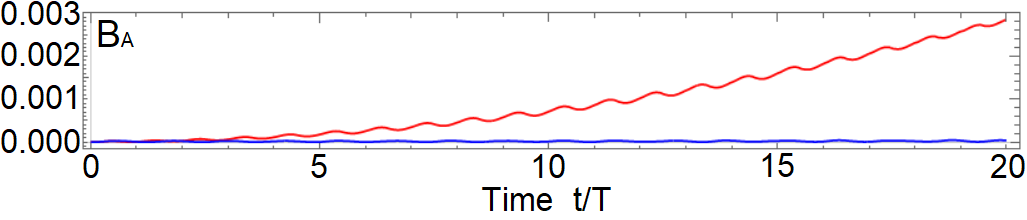}
	\caption[Time-evolution of the measure of backaction.]{Time-evolution of the measure of backaction $B_A$. Parameters are chosen such that $\delta\omega$ is extremal (blue) and $\delta\omega=0$ (red), with $\Omega/\sqrt{\omega}=0.005$.}
	\label{fig:backaction}
\end{figure}

 Its evaluation, together with the above considerations, leads us to the following conclusion: The backaction can be maximalized if $\cos\theta$ is maximal, that is, when $\delta\omega=0$, whereas for parameters which fulfill the $\cos\theta=0$ condition, the backaction on the excitation is minimal. This can be observed in Fig.(\ref{fig:backaction}), where the dominant feature (besides the continuous growth) is the $T$-periodic oscillation. We note that this oscillation largely corresponds to the periodical dynamics on phase space as described in \cite{G20}.

\subsection{Quantized excitation, quantized harmonics}\label{HHGQQ}

Incorporating the fully quantum nature of the dynamics is numerically  challenging without some kind of approximation. During our calculations, we employed the two-mode approximation, that is, considered only a single excitation and a single scattered mode. 
\\
A meaningful question, only treatable within the fully quantized formalism, is whether there are non-trivial correlations arising between absorption from the excitation mode and emission in the scattered modes.

\begin{figure}[h]
	\centering
	\includegraphics[width=1.03\linewidth]{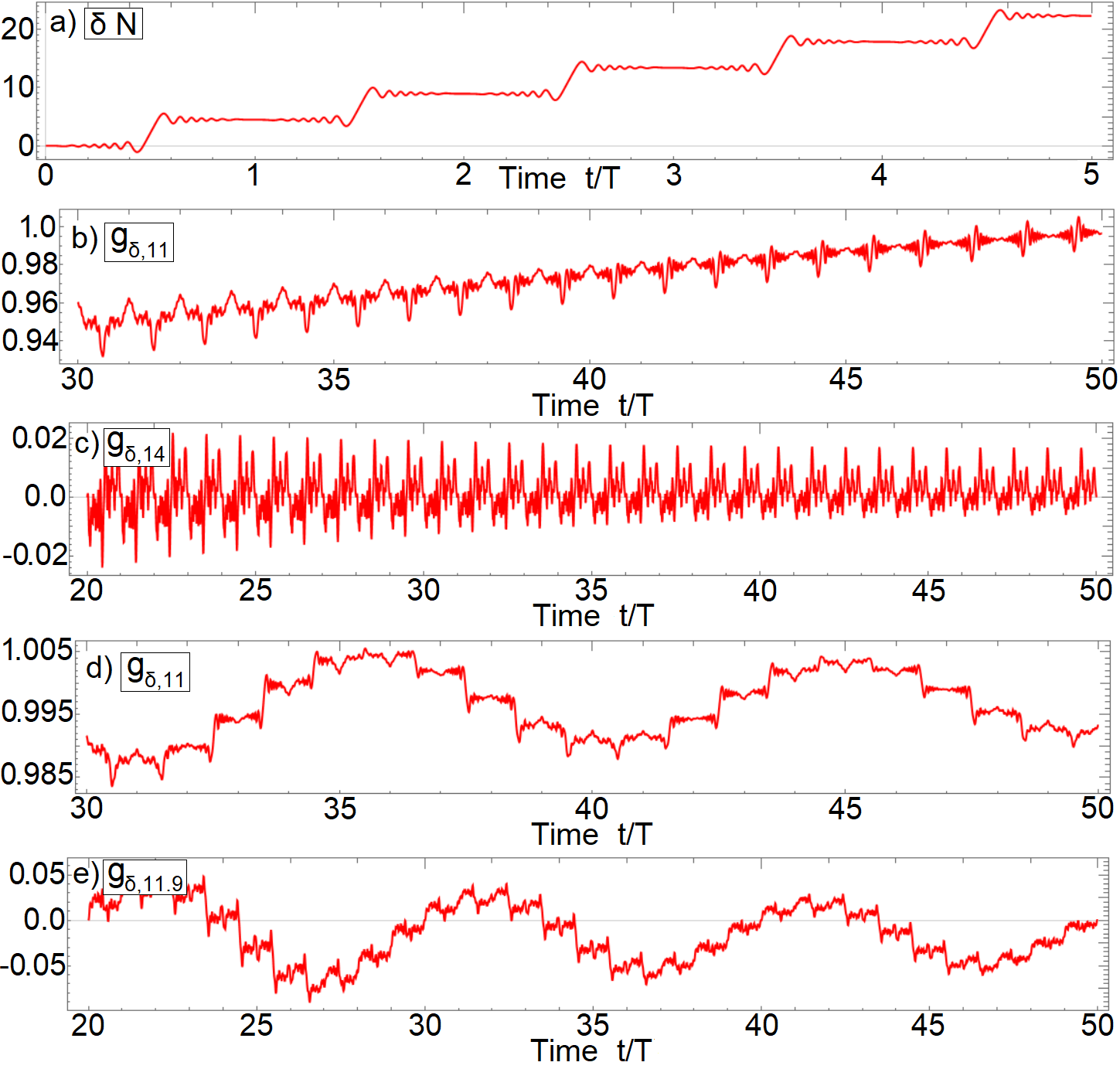}
	\caption[Correlations between excitation mode and harmonic mode.]{\textbf{a)} shows time-evolution of $\delta N$. Subfigures \textbf{b)} and \textbf{c)} shows the correlation between the absorption of a photon in the excitation mode and the emission of a photon in an odd- and even harmonic mode respectively, with the parameter chosen so that $|\delta\omega|=0$. Subfigures \textbf{d)} and \textbf{e)} are analogous, but with  $|\delta\omega|=0.1\omega$. }
	\label{fig:qqcorrelation}
\end{figure}

Let us introduce an operator measuring the number of absorbed photons in the excitation mode:
\begin{equation}
\delta N \equiv |\alpha^2|- N .
\end{equation} 
We define correlation function between $\delta N$ and $N_n$ as:
\begin{equation}
g_{\delta i} \equiv 
\dfrac{\langle \delta N N_i \rangle }{\langle \delta N \rangle \langle N_i \rangle},
\end{equation}
which, unlike previously introduced correlation functions, can be of negative value, since $\delta N$ can be negative.  The parameters have been chosen so that the monochromatic excitation contains $\approx 10^8$ photons.

In Fig.(\ref{fig:qqcorrelation}/a) we can observe that photon absorption from the highly populated excitation mode happens in discrete steps in each half-cycle.
There is a nearly unit correlation between the absorption from excitation mode and emission in odd-harmonic modes [see Fig.(\ref{fig:qqcorrelation})/b) and d)], however the even harmonic photon emissions are uncorrelated to the photon emission from the excitation [see Fig.(\ref{fig:qqcorrelation}/c) and /e)].

\section{Conclusions} \label{conclusion}
We analyzed photon statistics of high-order harmonics specific to a two-level radiating system. 
The harmonics induced by monochromatic excitations follow a relatively simple behaviour:
Odd harmonics oscillate between super-Poissonian and sub-Poissonian statistics, usually fulfilling the $Q\ll\langle N\rangle$ relation. Even harmonics (Hyper-Raman lines) can be, depending on the parameters, either strongly squeezed or effectively in the superposition of zero-, and one-photon states. 

Our results point to the theoretical possibility, that with specific excitations, HHG can be the source of one-photon radiation in many modes, encompassing a broad spectral range (notably with detuning, the spectra can become quasi-continuum) or a source of broadband squeezed states. 
\bigskip

Intermodal correlations within the radiation field have also been investigated. Generally speaking, the odd-odd harmonic photons are classically cross-correlated, while the  odd-, and even-harmonic photons are anticorrelated in all cases investigated by us. The even-even harmonic photons can be, depending on the parameters, either be strongly correlated or anti-correlated. Our results suggest that the anti-correlations correspond to nonclassical entanglement.

In other words, we have found that nonclassical properties, potentially of experimental interest, can be associated primarily with the modes of even-order harmonics.
HHG as a source of nonclassical light can be realized in the same experimental settings that allow observation of these optical lines, see E.g. \cite{BEBDPLMM19}.

\begin{acknowledgements}
This research was performed in the framework of the project GINOP-2.3.2-15-2016-00036. The project has also been supported by the European Union, co-financed by the European Social Fund, Grant No. EFOP-3.6.2-16-2017-00005—Ultrafast physical processes in atoms, molecules, nanostructures, and biological systems. Support by the ELI-ALPS project is also acknowledged. The ELI-ALPS project (GINOP-2.3.6-15-2015-00001) is supported by the European Union and co-financed by the European Regional Development Fund. We also acknowledge financial support from the Ministry of Innovation and Technology, Hungary Grant NKFIH-1279-2/2020.

\end{acknowledgements}

\appendix

\bigskip

\section{Analytical result for classically driven two-level atom}\label{Analy}

The spectrum of scattered radiation from a two-level system under monochromatic excitation is structured into qualitatively different odd- and even harmonics.
While the two-level system has been investigated in the literature thoroughly using semiclassical approach \cite{S65,GS92} --usually involving approximations that limit the validity of analytic results, or being given in a complicated form that offers little insight \cite{QXHW10}-- a transparent analytic characterization of the dynamics with respect to HHG, has not been given in the literature according to our knowledge.

Consider the semiclassical $H_{cc}$ Hamiltonian
\begin{equation}\label{H_c}
H_{cc}(t)= \dfrac{\hbar\omega_0}{2} \sigma_z
+ \dfrac{\hbar A}{2}\sigma_x \cos(\omega t \!+\! \phi_0),
\end{equation}
and use the unitary transformation:
\begin{align}\label{analunitarytr}
|\Psi'(t)\rangle
=
e^{\Lambda(t)} |\Psi(t)\rangle ,
\\
H'_{cc}(t)= 
e^{\Lambda(t)} H_{cc}  e^{-\Lambda(t)} + i\hbar e^{-\Lambda(t)} \dfrac{\partial}{\partial t} e^{\Lambda(t)} ,
\end{align}
with the choice 
\begin{equation}
\Lambda(t) \equiv i \dfrac{A}{2\omega}\xi \sin(\omega t \!+\! \phi_0) \sigma_x .
\end{equation}
Here $\xi \in \mathbf{R}$ is to be determined in the following.
The transformed Hamiltonian can be written as:
\begin{align}
H'_{cc}(t) =
\dfrac{\hbar\omega_0}{2} \bigg\{
\cos \bigg[ \dfrac{A}{\omega}\xi \sin(\omega t \!+\! \phi_0) \bigg]\sigma_z
\nonumber \\ 
+
\sin \bigg[ \dfrac{A}{\omega}\xi \sin(\omega t \!+\! \phi_0) \bigg]\sigma_y
\bigg\}
\nonumber \\ 
+ \dfrac{\hbar A}{2}(1-\xi)\cos(\omega t \!+\! \phi_0)\sigma_x .
\end{align}
Using the Anger-Jacobi identity, we can divide the Hamiltonian \cite{YLZ15} as:
$H'_{cc}(t)=H'_0 + H'_1(t) + H'_2(t)$, where the terms are the following:
\begin{widetext}
\begin{align}
H'_0 =
\dfrac{\hbar\omega_0}{2} J_0\bigg( \dfrac{A}{\omega}\xi\bigg) \sigma_z;
\\
H'_1(t) =
\dfrac{\hbar A}{2}(1-\xi)\cos(\omega t \!+\! \phi_0) \sigma_x
+\hbar\omega_0 J_1\bigg( \dfrac{A}{\omega}\xi\bigg)\sin(\omega t \!+\! \phi_0) \sigma_y;
\\
H'_2(t) =
\hbar\omega_0 \sum_{n=1}^{\infty} J_{2n}\bigg( \dfrac{A}{\omega}\xi\bigg)\cos[2n(\omega t \!+\! \phi_0) ] \sigma_z
+\hbar\omega_0 \sum_{n=1}^{\infty} J_{2n+1}\bigg( \dfrac{A}{\omega}\xi\bigg)\sin[(2n+1)(\omega t \!+\! \phi_0) ] \sigma_y.
\end{align}
\end{widetext}
With the neglection of $H_2'(t)$, a solution can be given \cite{LZZH12}. However, as  $H_2'(t)$ contains the terms associated with harmonic generations, we will need to incorporate it as the driving term in interaction picture.

At this point, let us fix $\xi$ such that:
\begin{equation}
J_1\bigg(\dfrac{A}{\omega}\xi \bigg)\omega_0
=
\dfrac{A}{2}(1-\xi)\equiv \dfrac{B}{4} .
\end{equation}
With this choice we can write:
\begin{align*}
H'_0 + H'_1(t) =
\\
\dfrac{\hbar\omega_0}{2} J_0\bigg( \dfrac{A}{\omega}\xi\bigg) \sigma_z
+ \dfrac{\hbar B}{4}\big( e^{-i(\omega t +\phi_0)}\sigma_+ 
+ e^{i(\omega t +\phi_0)}\sigma_- \big).
\end{align*}

The solutions can be found straightforwardly by applying the rotation transformation $e^{\tfrac{i}{2}(\omega t + \phi_0)\sigma_z}$, employing the $e^{\tfrac{i}{2}(\omega t + \phi_0)\sigma_z} \sigma_{\pm} 
e^{-\tfrac{i}{2}(\omega t + \phi_0)\sigma_z} 
\!=\! \sigma_{\pm} e^{\pm i (\omega t + \phi_0)} $ relation \cite{KC09}.
The transformed Hamiltonian and its eigenvalues turn out to be:
\begin{equation*}
\tilde{H}'_0 + \tilde{H}'_1 =\dfrac{\hbar}{2} \bigg[ \omega_0 J_0\bigg( \dfrac{A}{\omega}\xi\bigg) 
-\omega \bigg] \sigma_z
+ \dfrac{\hbar B}{4} \sigma_x
\end{equation*}
\begin{equation}\label{analiticeigenvalues}
\epsilon_\pm 
=
\pm \dfrac{\hbar}{2}\sqrt{ \big( J_0(\tfrac{A}{\omega}\xi)\omega_0 -\omega \big)^2 
	+ B^2/4  } .
\end{equation}
The eigenvectors are:
\begin{align}
| \tilde{e}\rangle = 
\sin\theta  |g \rangle
+
\cos\theta  |e \rangle ,
\nonumber \\
| \tilde{g}\rangle = 
\sin\theta  |e \rangle
-
\cos\theta  |g \rangle ,
\end{align}
where the $\theta$ parameter is given as:
\begin{equation}
\theta= 
\arctan\bigg[ \dfrac{\sqrt{ \big( J_0(\tfrac{A}{\omega}\xi)\omega_0 -\omega \big)^2 + B^2/4 } 
	-\left( J_0(\tfrac{A}{\omega}\xi)\omega_0 -\omega \right) }
{B/2} \bigg].
\end{equation}
The time-evolution can then be understood on the basis of eigenstates $| \tilde{e}\rangle$ and $| \tilde{g}\rangle$  in interaction picture. The driving is done by $\hbar W(t)\equiv e^{\tfrac{i}{2}(\omega t + \phi_0)\sigma_z} H'_2(t) 
e^{-\tfrac{i}{2}(\omega t + \phi_0)\sigma_z}$, where:

\begin{align}\label{Wdef}
W(t) =
\omega_0 \sum_{n=1}^{\infty} J_{2n}\bigg( \dfrac{A}{\omega}\xi\bigg)\cos[2n(\omega t \!+\! \phi_0)] \sigma_z 
\nonumber\\+
\omega_0 \sum_{n=1}^{\infty} J_{2n+1}\bigg( \dfrac{A}{\omega}\xi\bigg)\sin[(2n+1)(\omega t \!+\! \phi_0)] 
\nonumber\\ \times \big(\! \sin(\omega t \!+\! \phi_0 )\sigma_x + \cos(\omega t \!+\!\phi_0) \sigma_y \big).
\end{align}

The quantum state is written as:
\begin{align} \label{quantumstate}
|\Psi\rangle =
b^e(t) ~e^{i\tfrac{A\xi}{2\omega} \sin(\omega t + \phi_0 )\sigma_x} e^{\tfrac{i}{2}(\omega t+\phi_0)\sigma_z}  | \tilde{e} \rangle e^{-\tfrac{i}{\hbar}\epsilon_+ t}
\nonumber \\+
b^g(t) ~e^{i\tfrac{A\xi}{2\omega} \sin(\omega t + \phi_0)\sigma_x}
e^{\tfrac{i}{2}(\omega t+\phi_0)\sigma_z} | \tilde{g} \rangle e^{-\tfrac{i}{\hbar}\epsilon_- t}
\end{align}
with the time-dependence of coefficients, $b^e$ and $b^g$ given by:
\begin{align*}
i \dot{b}^e(t) = \langle \tilde{e} | W(t) | \tilde{e} \rangle b^e(t)
+ \langle \tilde{e} | W(t) | \tilde{g} \rangle e^{-i\tfrac{\epsilon_- - \epsilon_+}{\hbar}t} b^g(t) ,
\\
i \dot{b}^g(t) = \langle \tilde{g} | W(t) | \tilde{g} \rangle b^g(t)
+ \langle \tilde{g} | W(t) | \tilde{e} \rangle e^{-i\tfrac{\epsilon_+ - \epsilon_-}{\hbar}t} b^e(t) .
\end{align*}
The physical picture emerging is the following: The eigenstates of $\tilde{H}'_0 + \tilde{H}'_1(t)$ define two energy levels, which together with the unitary transform defines a set of infinite virtual energy levels (essentially equivalent to the Floquet quasi-energies). At the same time, $\hbar W(t)$ corresponds to higher-order optical processes and induces transitions between the eigenstates.

Using the following formulae:
\begin{widetext}
\begin{align*}
\langle \tilde{e} | \sigma_z | \tilde{e} \rangle
=
\cos^2\theta - \sin^2\theta
\\
\langle \tilde{g} | \sigma_z | \tilde{g} \rangle
=
\sin^2\theta - \cos^2\theta
\\
\langle \tilde{g} | \sigma_z | \tilde{e} \rangle =
\langle \tilde{e} | \sigma_z | \tilde{g} \rangle
=
2\sin\theta \cos\theta
\end{align*}
\begin{align*}
\langle \tilde{e} | \sigma_x | \tilde{e} \rangle
=
2\sin\theta \cos\theta 
\hspace{3.8cm}
\langle \tilde{e} | \sigma_y | \tilde{e} \rangle
= 0
\\
\langle \tilde{g} | \sigma_x | \tilde{g} \rangle
=
-2\sin\theta \cos\theta 
\hspace{3.8cm}
\langle \tilde{g} | \sigma_y | \tilde{g} \rangle
= 0
\\  
\langle \tilde{g} | \sigma_x | \tilde{e} \rangle
= \langle \tilde{e} | \sigma_x | \tilde{g} \rangle
=\sin^2\theta - \cos^2\theta
\hspace{1.3cm}
\langle \tilde{g} | \sigma_y | \tilde{e} \rangle
=-i
=- \langle \tilde{e} | \sigma_y | \tilde{g} \rangle
\end{align*}
the dynamical equations can be expanded, using the nonlinear optical parameter $\eta \equiv \tfrac{A\xi}{\omega}$ as below.
\begin{align}\label{analdynequ1} 
i \dot{b}^e(t) = b^e(t) \omega_0 \sum^{\infty}_{n=1}\bigg[ 
J_{2n}(\eta) \cos(2\theta) \cos[2n(\omega t + \phi_0)]
+ \dfrac{J_{2n+1}(\eta)}{2} \sin(2\theta) \big( \cos[2n(\omega t + \phi_0)] 
- \cos[(2n+2)(\omega t + \phi_0)] \big)
\bigg]
\nonumber\\ 
+  b^g(t) \omega_0 e^{-i\tfrac{\epsilon_- - \epsilon_+}{\hbar}t}
\sum_{n=1}^{\infty} \bigg[ 
J_{2n}(\eta) \sin(2\theta) \cos[2n(\omega t + \phi_0 )]
- \dfrac{J_{2n+1}(\eta)}{2} \cos(2\theta) \big( \cos[2n (\omega t + \phi_0)] 
- \cos[(2n+2)(\omega t + \phi_0)] \big)
\nonumber\\ 
+ i\dfrac{J_{2n+1}(\eta)}{2} \big( \sin[(2n+2)(\omega t + \phi_0 )] + \sin[2n(\omega t + \phi_0)]  \big) \bigg]
\end{align}
\begin{align}\label{analdynequ2} 
i \dot{b}^g(t) = -b^g(t) \omega_0 \sum^{\infty}_{n=1}\bigg[ 
J_{2n}(\eta) \cos(2\theta) \cos[2n(\omega t + \phi_0)]
+ \dfrac{J_{2n+1}(\eta)}{2} \sin(2\theta) \big( \cos[2n(\omega t + \phi_0)] 
- \cos[(2n+2)(\omega t + \phi_0)] \big)
\bigg]
\nonumber\\ 
+  b^e(t) \omega_0 e^{-i\tfrac{\epsilon_+ - \epsilon_-}{\hbar}t}
\sum_{n=1}^{\infty} \bigg[ 
J_{2n}(\eta) \sin(2\theta) \cos[2n(\omega t + \phi_0)]
- \dfrac{J_{2n+1}(\eta)}{2} \cos(2\theta) \big( \cos[2n(\omega t + \phi_0)] 
- \cos[(2n+2)(\omega t + \phi_0)] \big)
\nonumber\\ 
- i\dfrac{J_{2n+1}(\eta)}{2} \big( \sin[(2n+2)(\omega t + \phi_0)] 
+ \sin[2n(\omega t + \phi_0)]  \big) \bigg]
\end{align}
\end{widetext}

In semiclassical spectral calculations, the quantity of interest is $\langle D (t)\rangle\!\equiv\!\langle \Psi | d\sigma_x | \Psi \rangle$, the expectation value of the dipole-moment. 
\begin{align}
\langle D\rangle/d
=
\big(|b^+(t)|^2-|b^-(t)|^2 \big) \cos(\omega t + \phi_0 ) 
2\sin\theta \cos\theta 
\nonumber \\
+2\Re\big[ b^{+^*}(t) b^-(t) e^{i\tfrac{\epsilon_+ -\epsilon_-}{\hbar}t} \big]
\cos(\omega t + \phi_0 ) (\sin^2\theta-\cos^2\theta) 
\nonumber \\
+2\Im\big[ b^{+^*}(t) b^-(t)  e^{i\tfrac{\epsilon_+ -\epsilon_-}{\hbar}t} \big] 
\sin(\omega t + \phi_0) ,
\end{align}
Evaluation shows that the terms with not odd-harmonic frequencies have (plus-minus) $\delta\omega\equiv\tfrac{\epsilon_+-\epsilon_-}{\hbar}- \omega$ detuning from even-order multiples of the basic harmonic. 

\bigskip

\begin{figure}[h!]
	\centering
	\includegraphics[width=1.0\linewidth]{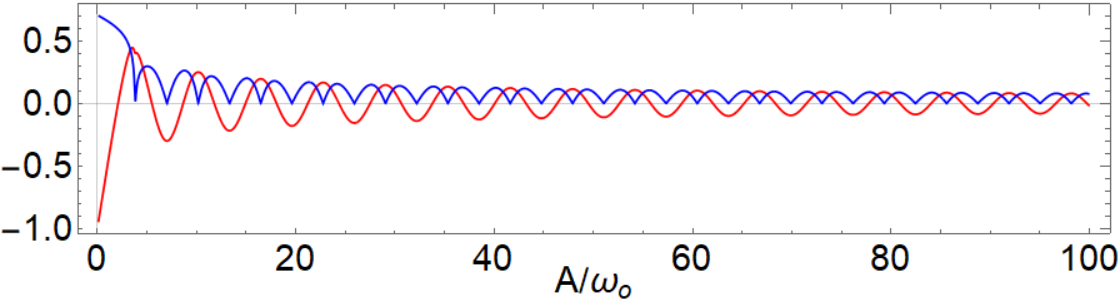}
	\caption[Dependence of $\delta\omega/\omega$ and $\cos\theta$.]{Dependence of $\delta\omega/\omega$ (red) and $\cos\theta$ (blue) on the amplitude of resonant excitation.}
	\label{fig:functions}
\end{figure}

If we fix the gap $\omega_0$, and detuning $\omega/\omega_0$ ratio, both  $\cos\theta$ and $\delta\omega$ are functions of only the amplitude, and are asymptotically (albeit with slow convergence) zero, see Fig.(\ref{fig:functions}).

Let us note that zero points of $\cos\theta$ are corresponding to local extremum of $\delta\omega/\omega$, that is, at these parameters the dual lines of even harmonics have maximal separation. The $\cos\theta$ function has zero points in all intensity range, more or less being distributed evenly.

\section{First-order perturbative expansion}\label{perturb}

\begin{figure}[h!]\hspace{-0.2cm}
	\centering
	\includegraphics[width=0.99\linewidth]{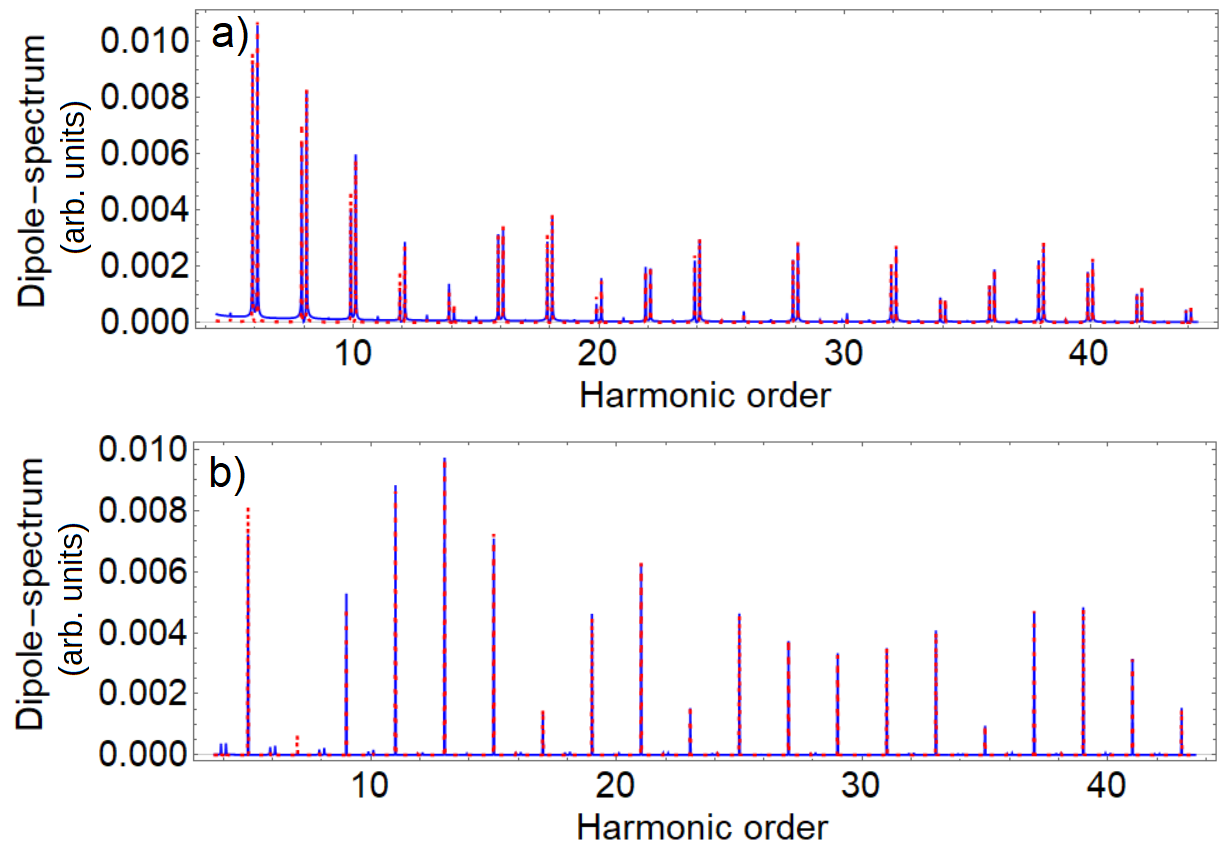}
	\caption[Comparison of numerical and perturbatively calculated spectra.]{Comparison of spectrum induced by resonant excitation, calculated numerically (blue) and with first-order perturbation method (dashed). Subfigure a) shows the spectra calculated with initial condition $b^e\!=\!b^g\!=\!1/\sqrt{2}$, and b) for initial condition $b^g\!=\!1$. }
	\label{fig:perturbative1}
\end{figure}

In the dynamical equations (\ref{analdynequ1}-\ref{analdynequ2}) there is no resonant contribution, that is, the $b^{(e/g)}$ coefficients follow high-frequency, small-amplitude oscillations around their initial values, which implies that perturbation methods are applicable. Comparison between spectra calculated numerically and through first-order perturbation --within realistic excitation intensity value-- can be seen on Fig.(\ref{fig:perturbative1}). We note that the dominant spectral lines (odd or even harmonics, depending on the initial conditions) are reproduced by the perturbative treatment typically within $\sim10\%$ relative error.

For the sake of simplicity, focus on the special case of $\cos\theta=0$, which, as mentioned above, corresponds to the maximal spectral gap between the dual lines of even-harmonics. Then equations (\ref{analdynequ1}-\ref{analdynequ2}) become:

\begin{widetext}
\begin{eqnarray*}
	i \dot{b}^e(t) = -b^e(t) \omega_0 \sum^{\infty}_{n=1}\bigg[ 
	J_{2n}(\eta)  \cos[2n(\omega t + \phi_0)] \bigg]
	\nonumber\\ 
	+  b^g(t) \omega_0 e^{-i\tfrac{\epsilon_- - \epsilon_+}{\hbar}t}
	\sum_{n=1}^{\infty} \bigg[ \dfrac{J_{2n+1}(\eta)}{2}  \big( \exp[2n i (\omega t + \phi_0)] 
	- \exp[-(2n+2)i(\omega t + \phi_0)] \big) \bigg]
\end{eqnarray*}
\begin{eqnarray*}
	i \dot{b}^g(t) = b^g(t) \omega_0 \sum^{\infty}_{n=1}\bigg[ 
	J_{2n}(\eta)  \cos[2n(\omega t + \phi_0)] \bigg]
	\nonumber\\ 
	+  b^e(t) \omega_0 e^{-i\tfrac{\epsilon_+ - \epsilon_-}{\hbar}t}
	\sum_{n=1}^{\infty} \bigg[ 
	\dfrac{J_{2n+1}(\eta)}{2}  \big( \exp[-2ni(\omega t + \phi_0)] 
	- \exp[(2n+2)i(\omega t + \phi_0)] \big)  \bigg]
\end{eqnarray*}
\end{widetext}

This special case is not unrealistic, considering that a careful selection of the parameters allows this condition to be fulfilled in all, not too specific intensity intervals.

Here we give the analytic expression of the first-order perturbation calculation results which have been employed in the article.
The first-order perturbative solution can be written as:
\begin{eqnarray}
b^e(t) \approx b^e(0) +i b^e(0) \zeta_1(t) - i b^g(0) \zeta_2(t)
\nonumber \\
b^g(t) \approx b^g(0) -i b^g(0) \zeta_1(t) - i b^e(0) \zeta^*_2(t)
\end{eqnarray}
where we define the $\zeta_1(t)$ and $\zeta_2(t)$ expressions as:
\begin{equation*}
\zeta_1(t) \!=\!
\omega_0  \sum^{\infty}_{n=1} \! \bigg[ 
\dfrac{J_{2n}(\eta)}{2n\omega} 
\big( \sin[2n (\omega t + \phi_0) ] - \sin[2n \phi_0 ] \big)
\bigg]
\end{equation*}

\begin{widetext}
\begin{align*}\hspace{-1.0cm}
\zeta_2(t) \!=\!
i\omega_0  \sum^{\infty}_{n=1} \! \left[ 
\dfrac{J_{2n+1}(\eta)}{2} 
\left(
\dfrac{\left[ 1 - e^{i(2n\omega + \tfrac{\epsilon_+-\epsilon_-}{\hbar})t} \right]}
{2n\omega + \tfrac{\epsilon_+-\epsilon_-}{\hbar}}e^{i2n\phi_0} 
+
\dfrac{\left[ 1 - e^{i(-(2n+2)\omega + \tfrac{\epsilon_+-\epsilon_-}{\hbar})t} \right]}
{(2n+2)\omega - \tfrac{\epsilon_+-\epsilon_-}{\hbar}}e^{-i(2n+2)\phi_0} 
\right)
\right]
\end{align*}
\end{widetext}
The dipole-operator expectation value can be expressed through $b^{e^*}(t)b^g(t) e^{i\tfrac{\epsilon_+-\epsilon_-}{\hbar}t}$ which term, after simplifications  can be rewritten as (\ref{bebg}).

The evaluation of the dipole-moment can be done in a lengthy but straightforward manner. The dipole-oscillation contains frequencies  $(2n+1)\omega$ and $(2n+1)\omega\pm \frac{\epsilon_+-\epsilon_-}{\hbar}=(2n+2)\omega\pm\delta \omega$. 

\begin{eqnarray}\label{bebg}
b^{e^*}(t)b^g(t) e^{i\tfrac{\epsilon_+-\epsilon_-}{\hbar}t}
=
\nonumber \\
b^{e^*}\!(0) b^{g}\!(0) 
\bigg[ 1 - i \zeta_1(t) \bigg]^2 e^{i\tfrac{\epsilon_+-\epsilon_-}{\hbar}t}
\nonumber \\ 
+ b^{g^*}\!(0) b^{e}(0) \bigg[\zeta^{*}_2(t) \bigg]^2
e^{i\tfrac{\epsilon_+-\epsilon_-}{\hbar}t}
\nonumber \\
+ \big( |b^{g}\!(0)|^2 - |b^{e}\!(0)|^2  \big)
\bigg[ i\zeta^*_2(t) + \zeta_1(t) \zeta^*_2(t) \bigg]
e^{i\tfrac{\epsilon_+-\epsilon_-}{\hbar}t}
\end{eqnarray}

If either $b^g(0)$ or $b^e(0)$ is zero, we can expect the lack of even-order harmonics in semiclassical solutions.
We note in passing, that the two spectral lines within even harmonics carry different multiples of the excitation phase, which can have consequences when macroscopic wave-propagation is considered.

\bibliography{allthebibGA}

\end{document}